\documentclass[prx,10pt,twocolumn,superscriptaddress,floatfix,showpacs]{revtex4-1}

\usepackage[british]{babel}  %Do hyphenation according to british english
\usepackage[scaled=0.86]{berasans}  % URL font that go well wtih palatino
\usepackage[colorlinks=true, allcolors=blue, urlcolor=blue]{hyperref}  %Hyperlinks (pink, green, blue)
\usepackage{graphicx} % Package to insert exteral figures
\usepackage[babel]{microtype}  %Improves text justification
\usepackage{amsmath,amssymb,amsthm,bm,amsfonts,mathrsfs,bbm} %Usefull math packages

\usepackage{xspace}  %Useful to add space in macros
\usepackage{pgfplots}
\usepackage{xcolor,colortbl}
\usepackage{array}
\usepackage{bigstrut}

\newcommand{\tr}{\text{tr}}
\newcommand{\ket}[1]{| #1 \rangle}
\newcommand{\bra}[1]{\langle #1|}

\newcommand{\ip}[2]{\langle #1|#2 \rangle}

\newcommand{\be}{\begin{equation}}
\newcommand{\ee}{\end{equation}}
\newcommand{\bea}{\begin{eqnarray}}
\newcommand{\eea}{\end{eqnarray}}
\newcommand{\bes}{\begin{equation*}}
\newcommand{\ees}{\end{equation*}}
\newcommand{\beas}{\begin{eqnarray*}}
	\newcommand{\eeas}{\end{eqnarray*}}

\newcommand{\x}{\mathrm{x}}
\newcommand{\ketbra}[1]{\ket{#1}\!\bra{#1}}

\def\x{\mathrm{x}}
\def\y{\mathrm{y}}
\def\g{\mathrm{guess}}

\def\tr{\mathrm{tr}}
\def\I{\mathbb{I}}

\def\Q{\mathrm{Q}}
\def\NC{\mathrm{NC}}

\newtheorem*{thm*}{Theorem}

\newtheorem*{lem*}{Lemma}

\newtheorem*{lipschitzLem*}{Lemma \ref{lipschitz}}
\newtheorem*{lipschitzCubeLem*}{Lemma \ref{lipschitzCube}}
\newtheorem*{pgmNearlyOptimalThm*}{Theorem \ref{pgmNearlyOptimal}}

\begin{document}

%\title{  Contextual advantages for certifiable maximum confidence discrimination}

%\title{Contextual Advantages for State Discrimination and Its Certification}

%\title{Maximum confidence discrimination: contextual advantages and certification}

%\title{Contextual Advantages for State Discrimination Can Be Certified }

%\title{Contextual advantages and certification for maximum confidence discrimination}

\title{Contextual advantages and certification for maximum confidence discrimination}

%#####################################

\author{ Kieran Flatt}
\affiliation{School of Electrical Engineering, Korea Advanced Institute of Science and Technology (KAIST), 291 Daehak-ro, Yuseong-gu, Daejeon 34141, Republic of Korea }

\author{Hanwool Lee }
\affiliation{School of Electrical Engineering, Korea Advanced Institute of Science and Technology (KAIST), 291 Daehak-ro, Yuseong-gu, Daejeon 34141, Republic of Korea }

\author{ Carles Roch i Carceller}
\affiliation{Department of Physics, Technical University of Denmark, 2800 Kongens Lyngby, Denmark}

\author{Jonatan Bohr Brask}
\affiliation{Department of Physics, Technical University of Denmark, 2800 Kongens Lyngby, Denmark}

\author{Joonwoo Bae}
%\email{joonwoo.bae@kaist.ac.kr}
\affiliation{School of Electrical Engineering, Korea Advanced Institute of Science and Technology (KAIST), 291 Daehak-ro, Yuseong-gu, Daejeon 34141, Republic of Korea }

%########################################

\begin{abstract}
%One of the most fundamental results in quantum information theory is that no measurement can perfectly discriminate between non-orthogonal quantum states. In this work, we investigate quantum advantages for discrimination tasks over classical theories by considering a maximum confidence measurement that unifies different strategies of quantum state discrimination, including minimum-error or unambiguous discrimination. We first show that maximum confidence discrimination, as well as unambiguous discrimination, contains contextual advantages. We then consider a semi-device independent scenario of certifying a maximum confidence measurement, where a measurement apparatus is not fully trusted. The scenario naturally consists of undetected events, since a maximum confidence measurement takes detected events only into account. We show that the certified maximum confidence in quantum theory also contains contextual advantages. Our results establish how the advantages of quantum theory over a classical model may appear in a realistic scenario of a discrimination task. 

One of the most fundamental results in quantum information theory is that no measurement can perfectly discriminate between non-orthogonal quantum states. In this work, we investigate quantum advantages for discrimination tasks over noncontextual theories by considering a maximum confidence measurement that unifies different strategies of quantum state discrimination, including minimum-error and unambiguous discrimination. We first show that maximum confidence discrimination, as well as unambiguous discrimination, contains contextual advantages. We then consider a semi-device independent scenario of certifying maximum confidence measurement. The scenario naturally contains undetected events, making it a natural setting to explore maximum confidence measurements. We show that the certified maximum confidence in quantum theory also contains contextual advantages. Our results establish how the advantages of quantum theory over a classical model may appear in a realistic scenario of a discrimination task. 
 \end{abstract}

%\pacs{03.65.Ud, 02.50.Le, 03.67.Ac}

\maketitle

\section{Introduction}

Quantum information processing displays advantages over its classical counterpart. These gaps have their origins in fundamental results that show how the two types of theories differ. Quantum key distribution protocols, for example, exploit the indistinguishability of non-orthogonal states to establish security without any assumptions on the computational capabilities of adversaries \cite{PhysRevLett.68.3121}. Likewise, measurements on entangled states may give rise to nonlocal correlations, which cannot be produced from classical systems \cite{PhysicsPhysiqueFizika.1.195, RevModPhys.86.419}. Nonlocal correlations lead to various practical quantum information applications, in particular device-independent quantum information processing, including secure communication \cite{ekert1991, PhysRevLett.98.230501, Pironio_2009} and randomness generation \cite{Pironio:2010tu, Acin:2016tk}. In addition, nonlocal correlations can be exploited for the certification of quantum resources such as entanglement, which enables the aforementioned advantages for quantum information processing \cite{PhysRevLett.121.180503}. 

The fact that two nonorthogonal states cannot be perfectly discriminated is among the most fundamental results in quantum information theory \cite{Helstrom:1969aa}. This is closely connected to other key results, such as the quantum no-cloning theorem \cite{Wootters:1982aa} and no-signaling condition \cite{GISIN19981}. If perfect clones of a pair of non-orthogonal states could be obtained, it would be possible to perfectly discriminate the states. Conversely, perfect discrimination between non-orthogonal states makes it possible to prepare copies of the states. Quantum cloning converges, in the asymptotic limit, to quantum state discrimination \cite{bae2006}, which is then limited by the no-signaling condition \cite{PhysRevA.71.062315, PhysRevLett.107.170403}. The results for a pair of non-orthogonal states have been applied to quantum cryptographic protocols \cite{bennett1992} and various other tasks in quantum information theory \cite{chefles2000, bergou2004, bergou2007, barnett2009, bae2015}. 

In this work, we compare the limits of quantum state discrimination with those of classical physics, in the sense of noncontextual theories. The distinction between the two types of theory in the task of two-state discrimination has recently been shown \cite{PhysRevX.8.011015}. In a noncontextual ontological model, operationally equivalent experimental procedures have the same representation. This feature does not hold for quantum theories, so noncontextuality can be understood as one form of classicality. In the aforementioned work, the maximal success probability in two-state, minimum-error discrimination (MED) is characterised in a noncontextual ontological model. It turns out that two-state MED in quantum theory is more successful than the derived limitation, showing contextual advantages for quantum state discrimination. 

%Interestingly, the limitations can be isomorphically related to the characterisation of local probabilities with binary outcomes. Thus, nonlocal advantages may also be concluded. Note also that the no-signaling condition in a Bell scenario is transferred to the operational equivalence in MED. Very recently, similar to the relations between quantum state discrimination and quantum cloning, the contextual advantages for state-dependent quantum cloning were shown \cite{Lostaglio2020contextualadvantage}. 

From the point of view of realising these quantum advantages, a general difficulty lies in the inherent noise of quantum measurements. Even if a state has been prepared, it will sometimes not be detected due to, for example, photon losses. This is treated in MED by binning such cases among the possible outcomes at the cost of increasing the error rate. 

Another form of quantum state discrimination may be considered. In unambiguous discrimination (UD), a conclusion from certain detection events is never wrong but there is a possibility that no guess is returned \cite{DIEKS1988303, Ivanovic:1987aa, PERES198819}. An additional arm that collects all inconclusive outcomes is included. The possibility of realising UD, however, highly depends on parameters such as the Hilbert space dimension and the number of states. For instance, for qubit states it cannot be realised for cases other than two pure states.  

A figure of merit that operationally unifies the different senses of quantum state discrimination is the confidence \cite{PhysRevLett.96.070401}. The confidence is defined as the probability that, given a detection event, a detector correctly concludes that a state, chosen among an ensemble, has been prepared. In a maximum confidence measurement this figure of merit is maximised. Detectors in UD have certainty as the maximum confidence since a detection event never leads to a wrong conclusion. A maximum confidence measurement (MCM) performs MED if the confidence over the whole ensemble of states is considered. 

It should be noted that MCMs are concerned with detected events only. The consequence is that MCMs do not suffer from the same weaknesses as MED or UD. This is closely connected to a retrodictive view of quantum theory, whereby detected events in the present assert statements about state preparation in the past, as discussed in a recent review \cite{sym13040586}. One may therefore exploit MCMs to pave a way to gain contextual advantages with imperfect measurement devices in a realistic setting. It is also possible, taking a different point of view of retrodictive quantum theory, to certify the maximum confidence one can have in uncharacterised detectors used for state discrimination. This may be interpreted as a semi-device-independent scenario, \cite{VanHimbeeck2017semidevice, PhysRevLett.126.210503} under the assumption that states are well-characterised but the measurements not at all.

Here, we establish contextual advantages for both state discrimination and its certification in a realistic scenario where undetected events may appear. We first present contextual advantages for UD by showing that the minimal rate of inconclusive outcomes in quantum theory is strictly lower than that in a noncontextual ontological model. Then, the contextual advantages are shown for maximum confidence discrimination: an MCM in quantum theory gives rise to a higher maximum confidence over a noncontextual theory. We next consider a semi-device-independent scenario with uncharacterised detectors. We develop the framework of certifying the maximum confidence in the scenario given a preparation of states and detected events. It is shown that the certifiable maximum confidence in quantum theory contains contextual advantages in the realistic scenario that may contain undetected events. Our results provide the unifying framework for the existence and the certification of contextual advantages in a realistic quantum state discrimination scenario. 

%In the certification of an MCM, it is found that the more frequently detection events occur in a single detector, the lower the the maximum confidence will be. Conversely, as the maximum confidence on a detector increases, the rate of detection events lowers. A single detector performing UD, where the maximum confidence is the unit, always shows a lower rate of detection events than a detector used in MED. We reiterate that the certification of maximum confidence can be performed with detected events only. 

The paper is organised as follows. In Sec. \ref{section:bg}, we begin with a summary of different figures of merits in quantum state discrimination. The contextual advantages for minimum error, unambiguous and maximum confidence quantum state discrimination are then shown in Sec. \ref{section:con}. We then present the certification of an MCM in Sec. \ref{section:cert}. Two-input and three-outcome scenarios, with one arm containing the undetected events only, are considered. In Sec. \ref{section:adv}, we compare quantum and noncontextual theories in the certification of an MCM, then include noise in our model in Sec. \ref{section:noisyadv}. Finally, we summarise the results and discuss related questions in Sec. \ref{section:conc}.

\section{  Background } \label{section:bg}

%\section{ Quantum state discrimination}

Let us begin by collecting the terminology and notation to be used throughout the manuscript. We also summarise different figures of merits in quantum state discrimination. 

For convenience, state discrimination can be framed as a communication protocol for two parties, named Alice for preparation and Bob for measurement. Alice prepares her quantum system in one of the states in  an ensemble of $n$ states, denoted by 
\bea
\mathrm{ensemble}:~S_n = \{q_{\x}, \rho_{\x} \}_{\x=1}^n,  \label{eq:ens}
\eea 
which describes a state $\rho_{\x}$ is generated with {\it a priori} probability $q_{\x}$ for $\x=1,\cdots, n$. Bob then performs an $n$ outcome measurement described by positive-operator-valued-measure (POVM) elements
\bea
\mathrm{measurement}:~M= \{ M_{\y}\}_{\y=1}^n,  \label{eq:mea}
\eea 
each of which may be optimised to give a correct guess about a state that has been prepared. For completeness, the condition $\sum_{\y}M_{\y} = \I$ must be satisfied. 

\subsection{ Minimum error discrimination and unambiguous discrimination }

In MED, the figure of merit, called the guessing probability $P_{\g}$, is the highest probability of guessing $x$ correctly on average:
\bea
P_{\g} = \max \sum_{\x=1}^n q_{\x} \tr[\rho_{\x} M_{\x}],  \label{eq:med}
\eea
where the maximisation runs over a complete measurement. A closed form of the maximal success probability is known for two states in general,
\bea
P_{\g} = \frac{1}{2} + \frac{1}{2} \| q_1\rho_1 -q_2 \rho_2 \|_1, \label{eq:pguess}
\eea
where $\|  X \|_1 = \tr\sqrt{X^{\dagger} X}$. Otherwise, a closed form has been found in some specific cases only \cite{bae2013, PhysRevA.87.012334, PhysRevA.87.062302, Weir_2018}. While the error, averaged over the states in $S_n$, is minimised, not all detection events lead to a correct guess. Undetected events are binned in randomly chosen outcomes, which will make the guessing probability unattainable.

Detection events in UD measurements identify states with certainty. This is possible if the probability of outcome $\y$ given a state $\rho_{\x}$ is given by
\bea
P_{\mathrm{M|P}} (\y | \x) := \tr[\rho_{\x} M_{\y}] \propto \delta_{\x,\y}. \label{eq:ud}
\eea
where $\mathrm{M}$ and  $\mathrm{P}$ denote a measurement and a preparation, respectively. This shows that the detector described by $M_{\y}$ responds to $\rho_{\x}$ but not the other states. Under this condition, it may, however, appear that a measurement is not complete, i.e., $\sum_{\y} M_{\y} < \I$. An additional outcome $M_0$ is included to fulfill the completeness condition:
\bea
\sum_{\y=0}^n M_{\y} =\I. 
\eea 
The arm described by $M_0$ collects those detection events which give ambiguous conclusions. Then, in the case of UD, a conclusion from a detection event is completely unambiguous since no error in the legitimate arms is permitted. For qubit states, this is possible only when two pure states can be prepared. Preparation of pure states with certainty would not be feasible in a realistic setting either. We can say that it is not practical to meet the conditions in Eqs. (\ref{eq:pguess}) and (\ref{eq:ud}) in MED and UD, respectively.

\subsection{ Maximum confidence discrimination }

The notion of confidence for a detection event in a discrimination task has been defined as the conditional probability
\cite{PhysRevLett.96.070401}:
\bea
\mathrm{confidence:}~~ C(\y) = P_{\mathrm{P|M}} ( \y | \y), 
\eea
i.e., the probability that a detection event corresponding to $M_{\y}$ correctly indicates that a preparation was $\rho_{\y}$. One can interpret this retrodictively, as a detected event implying a conclusion about state preparation done in the past. 

The confidence may be computed with quantum probabilities by using Bayes' rule, 
\bea
C^{(\Q)}(\y)= \frac{ P_{\mathrm{P}} ( \y) P_{ {\mathrm{M|P}}} (\y | \y) }{ P_{\mathrm{M}} (\y)} =   \frac{q_{\y} \tr[M_{\y} \rho_{\y} ]}{ \tr[ M_{\y} \rho]},  ~~~~~ \label{eq:defexp}
\eea
where $P_{\mathrm{M}}(\y)$ is the probability of a detection event on the detector $M_{\y}$ for an ensemble $\rho  = \sum_{\x} q_{\x} \rho_{\x}$ and $P_{\mathrm{P}}(\y) = q_{\y}$ the {\it a priori} probability. Hence, an MCM aims to maximises the confidence above. Throughout, an MCM in quantum theory is denoted by
\bea
\max ~C^{(\Q )} ({\y} )
\eea
where the maximisation runs over all measurements. Note that an MCM can be defined for any ensemble in Eq. (\ref{eq:ens}).

We remark that maximum confidence discrimination is well-fitted to a realistic scenario including imperfect preparations and measurements. Firstly, it can be adapted to cases where the detected measurement statistics are not complete whereas MED can only find the optimal guessing probability whenever a measurement is complete. As MCM is concerned with detected events only, undetected ones can be counted as ambiguous outcomes. Secondly, an MCM can be considered for ensembles for which unambiguous measurement outcomes can not be obtained. MCM presents, for these reasons, a more realistic setting for identifying a state among a given ensemble. 

Maximum confidence discrimination also provides a unifying framework of the aforementioned figures of merits in state discrimination. An MCM coincides with UD if $C(\y)=1$ for all $\y$. In this sense, whenever UD is possible for an ensemble, it will emerge as the MCM. One can also apply an MCM to maximise the success probability over an ensemble or a subensemble by taking into account in the possibility of undetected events occuring:
\bea
\max \sum_{\y} p_M(\y) C(\y) = \max  \sum_{\y} q_{\y} \tr[M_{\y}\rho_{\y}] \label{eq:ba}
\eea
where the maximisation runs over a complete measurement. An MCM as defined above reproduces MED if the inconclusive outcome rate is zero. It is also worth noting that optimal measurements for MED, UD, and maximum confidence discrimination are generally not identical \cite{PhysRevLett.96.070401}.

\section{Contextual advantages for quantum state discrimination} 
\label{section:con}

Finding circumstances in which quantum experiments perform differently to their classical equivalents is central to the field of quantum information theory. State discrimination is a fundamental task in many practical applications. It is natural to ask how it differs between quantum and classical theories. This was recently addressed in Ref. \cite{PhysRevX.8.011015}, where it was shown that MED of quantum states contains contextual advantages. 

In this section, we consider UD and MCM and show contextual advantages. For the latter case, a pair of mixed states for which UD cannot be achieved are considered. Thus, we show contextual advantages for quantum state discrimination in general. We begin with a review of noncontextual ontological models and then consider MED, UD and MCM. 

\subsection{Noncontextual ontological model}

%One method for doing this is to determine whether or not a given scenario is contextual \cite{PhysRevA.71.052108, PhysRevX.8.011015, schmid2020structure}. Roughly speaking, the probability that a contextual event occurs depends upon the alternative events which could, but do not, occur. This definition can be made more precise with the use of an ontological model.

\begin{figure*}[]
    \centering
    \includegraphics[scale=0.75]{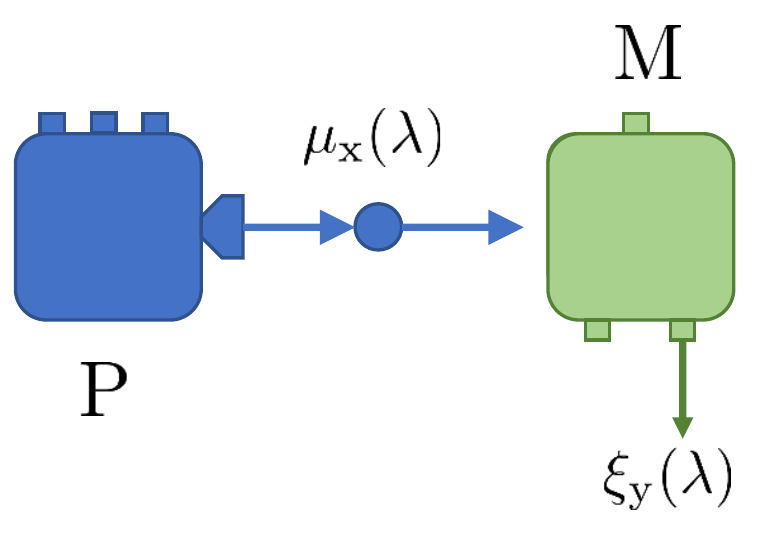} \hspace{1cm} \includegraphics[scale=0.7]{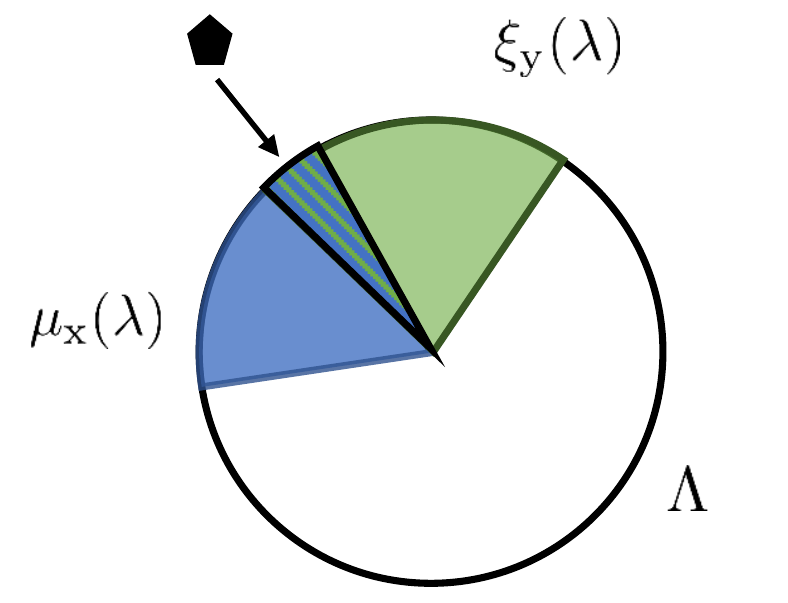}
    \caption{ A prepare-and-measure scenario in an ontological model (left) is modelled by the ontic state space $\Lambda$ (right). Note that $\mathrm{P}$ denotes a preparation represented by an epistemic state $\mu_{\x}(\lambda)$ and $\mathrm{M}$ a measurement by a response function $\xi_{\y} (\lambda)$. The probabilities extracted from the theory are given by integrals over the overlap marked by a black pentagon, see Eq. (\ref{eq:ossprob}).  }
    \label{fig:ontspace}
\end{figure*}

An operational theory contains descriptions of possible operations, such as preparations and measurements, and a prescription for calculating probabilities of measurement outcomes. Let us here review noncontextual ontological models \cite{PhysRevA.71.052108, PhysRevX.8.011015, schmid2020structure} and characterise preparation noncontextuality.

Let $\Omega$ denote an ontic state space so that an element $\lambda \in \Omega$ fully characterises the physical properties of a given system. A state preparation $\x$ corresponds to a sample of the ontic state space up to a probability distribution $\mu_{\x} (\lambda)$, which is called an epistemic state. A measurement $M$ contains a set of possible outcomes that occur with a dependence on the ontic state space. An outcome denoted by $\y$ is represented by a response function $\xi_{\y | M} (\lambda)$ that satisfies
\bea
\mathrm{positivity:}&~&\xi_{\y |M} (\lambda) \geq 0,~\forall \y,~\forall\lambda   ~\mathrm{and} \nonumber \\
\mathrm{completeness:}&~&  \sum_{\y} \xi_{\y|M} (\lambda) = 1,~\forall \lambda. \label{eq:ncon}
\eea
so that it can be interpreted as a probability distribution over the ontic states. Probabilities extracted from the ontological model with a preparation $\x$ and a measurement $M$ are given by 
\bea
{\rm P}( \y | \x, M) = \int_{\Omega} ~d \lambda ~ \mu_{\x}(\lambda) \xi_{\y | M}  (\lambda). 
\eea

\par

The preparation noncontextuality criterion is then identified as follows. Consider two preparations $\mu_{\x} (\lambda)$ and $\mu_{\x^{\prime}} (\lambda)$ that cannot be distinguished by any measurement, i.e., no response function provides different probabilities for the preparations. These preparations are called \textit{operationally equivalent}. A model is then {\it preparation noncontextual} if the operational equivalence of a set of preparatory processes implies that they are represented by the same epistemic state:
\bea \label{eq:ossprob}
&&{\rm P}( \y | \x, M) = {\rm P}( \y | \x^{\prime},  M) \, \, \, \forall \{ \y | M \} \nonumber\\
&& \implies \mu_{\x}(\lambda) = \mu_{\x^{\prime}}(\lambda) \forall \lambda. 
\eea
Measurement noncontextuality can also be defined in a similar manner. %We are here are concerned with preparation noncontextuality only. In fact, the former can be shown to follow if the latter is assumed \cite{PhysRevX.8.011015}. 

%The probabilistic predictions of a model are constrained if noncontextuality is imposed and the model cannot recreate quantum theory. As a result, a number of researchers have found examples of quantum information processes that are more powerful than their classical analogues. Of particular interest is a recent article concerning noncontextual inequalities for MED. The authors assume measurement noncontextuality and then calculate the maximum value taken by the relevant figure of merit. It is less than the maximum success probability in the equivalent quantum experiment. A similar inequality has subsequently been found for cloning \cite{Lostaglio2020contextualadvantage} and a more general algorithm for deriving inequalities has been presented \cite{PhysRevA.97.062103}.

Having introduced an operational framework above, we use definitions and notations in the following manner. For an epistemic state $\mu_{\x}(\lambda)$, a non-overlapping state is denoted by $\overline{\mu}_{\x} (\lambda)$ such that 
\bea
\mu_{\x}(\lambda)\overline{\mu}_{\x}(\lambda)=0,~~ \forall \lambda\in \Omega.
\eea
%In preparation noncontextual theories, an epistemic state  $\mu_{\x}(\lambda)$ uniquely defines its non-overlapping state $\overline{\mu}_{\x}(\lambda)$. This does not necessarily hold for other theories. 
The support of an epistemic state $\mu_{\x}$ is defined as
\bea
{\rm supp} [ \mu_{\x}(\lambda)] = \{ \lambda\in \Omega~:~ \mu_{\x}  (\lambda) \neq 0 \}. 
\eea
For instance, we have ${\rm supp} [ \mu_{\x}(\lambda)] \cap {\rm supp} [ \overline{\mu}_{\x}(\lambda)] = \emptyset$.

% the region of $\Omega$ for which $\mu_{\x}  (\lambda) \neq 0 $. 

An important set of response functions is the set representing projectors from quantum theory. In quantum theory, each POVM element $E_{\y}$ of a projective measurement satisfies $ {\rm tr} [E_{\y} |\x \rangle \langle \x |] = \delta_{\x,\y}$ for an ensemble $\rho_{\x}$ for $\x=1,...,N$ which form a basis. In an operational theory, $E_{\y}$ is represented by $\xi_{\y}(\lambda)$ and $\rho_{\x}$ by $\mu_{\x}(\lambda)$. It has been shown that, in noncontextual theories, the corresponding response functions take the form  \cite{PhysRevX.8.011015}
\begin{equation} \label{eq:xinc}
\xi_{\y}(\lambda) = \begin{cases} 1 & \mbox{if } \lambda \in {\rm supp} [\mu_{\y} (\lambda)] \\ 0 & \rm{otherwise} \end{cases}
\end{equation}
i.e., they are \textit{outcome deterministic}. 

For two-state discrimination in a noncontextual model, a useful quantity is the confusability, which is the probability of finding the outcome $x$ given a measurement on a different state $\mu_{\y} (\lambda)$ \cite{PhysRevX.8.011015, PhysRevLett.110.120401}. In a preparation noncontextual model, the confusability for a pair of states $\mu_{\x}(\lambda)$ and $\mu_{\y}(\lambda)$ can be defined as follows
\bea
c_{\x,\y} = \int_{ {\rm supp}[\mu_{\x}(\lambda)]} d\lambda \mu_{\y}(\lambda). \label{con}
\eea
In quantum theory, the confusability for two pure states can be identified with the state overlap 
\bea
c_{\x,\y}  = \tr[ |\psi_{\x}\rangle \langle \psi_{\x}| ~|\psi_{\y}\rangle \langle \psi_{\y}| ]  = |\langle \psi_{\x} | \psi_{\y} \rangle|^2. \label{eq:confusability}
\eea
It is clear that that the confusability is symmetric, i.e., $c_{\x,\y} = c_{\y,\x}$.

\subsection{ Contextual advantages for MED }

In Ref. \cite{PhysRevX.8.011015}, MED for two states in a noncontextual model was considered, and contextual advantages for MED of quantum states were shown. %The confusability characterises two states considered in state discrimination. 

%It is useful to characterise the discrepancies between contextual and noncontextual theories in terms of specific scenarios. A particularly strong demonstration of this behaviour has been made for MED. It is worth discussing the derivation of this result \cite{PhysRevX.8.011015} in more detail.

Suppose that two quantum states $|\psi_1 \rangle $ and $|\psi_2 \rangle$ are provided, for which an optimal measurement for MED is denoted by $M=\{M_{1}, M_{2} \}$. Two states can be characterised by the angle between them,
\bea
\cos \theta = \langle \psi_1 | \psi_2 \rangle = \sqrt{c_{1,2}} \label{eq:c}
\eea
where $c_{1,2}$ is the confusability: it suffices to consider a two dimensional Hilbert space. It is clear that one can find the statistics of measurement outcomes from the states and the measurement. The guessing probability for two quantum states in Eq. (\ref{eq:pguess}) can be straightforwardly computed. 

The ensemble consisting of the states $|\psi_1\rangle$ and $|\psi_2\rangle$ only, however, does not imply any equivalence relations so that noncontextuality cannot yet be used to constrain the model. Another pair of states, $|\overline{\psi}_{1} \rangle $ and $|\overline{\psi}_2 \rangle$,  must be used. The overlap and optimal guessing probability of this ensemble are identical to those of the former. The two pairs of states are related by
\bea
\frac{1}{2} (|\psi_1 \rangle\langle \psi_1| + |\overline{\psi}_1 \rangle\langle \overline{\psi}_1|) = \frac{1}{2}( |\psi_2 \rangle\langle \psi_2| + |\overline{\psi}_2  \rangle\langle \overline{\psi}_2|) = \frac{\I}{2} .~~~ ~~\label{eq:rel}
\eea
This provides an equivalence relation between the two quantum ensembles which can be used to derive relations between epistemic states.

A noncontextual model is then constructed such that it is consistent with this equivalence relation. Two epistemic states, denoted by $\mu_1 (\lambda)$ and $\mu_2(\lambda)$, can be introduced so that they have the same confusability $c_{1,2}$ with the quantum states Eq. (\ref{eq:c}), see also Eqs. (\ref{con}) and (\ref{eq:confusability}). The state space in a noncontextual model should also satisfy the equivalence relation. This implies that there exist mirrored states $\overline{\mu}_1(\lambda)$ and $\overline{\mu}_2(\lambda)$ such that 
\bea
\frac{1}{2} \mu_1 (\lambda) + \frac{1}{2} \overline{\mu}_1(\lambda) = \frac{1}{2} \mu_2(\lambda) + \frac{1}{2} \overline{\mu}_2(\lambda) = \frac{\mu_{\I/2} (\lambda)}{2} , 
\eea
consistent with Eq. (\ref{eq:rel}). Note also that the mirrored states share the same confusability with the original pair.  

In Ref. \cite{PhysRevX.8.011015}, it is shown that the preparation nontextuality constrains the statistics in terms of various sharp measurements, see Eq. (\ref{eq:xinc}) and finds the guessing probability as follows, 
\bea
P_{\g}^{\rm (NC)}  =1 - \frac{1}{2} c_{1,2} 
\eea
which is strictly less than the quantum bound in Eq. (\ref{eq:pguess}), i.e., 
\bea
P_{\g}^{\rm (Q)}  =\frac{1}{2} + \frac{1}{2} \sqrt{ 1- c_{1,2}}. 
\eea
This result, known as the Helstrom bound, is significant in that it shows that the predictions of noncontextual theories differ quantitatively from those of quantum theory. The results can also apply to mixed states when noise is present. 
 
\subsection{ Contextual advantages for UD }
\label{subsectionIIC}
 
Another scenario in state discrimination is UD, where, rather than finding the highest success probability over an ensemble, each state is identified with certainty. As with MED, a noncontextual model of UD can be constructed. Given the constraint of UD, the aim is to minimise the probability of having inconclusive outcomes. In what follows, we define UD for noncontextual theories and derive a noncontextual inequality associated with the rate of inconclusive outcomes, from which contextual advantages for quantum UD are shown. 

\begin{figure}[]
    \centering
    \includegraphics[scale=0.24]{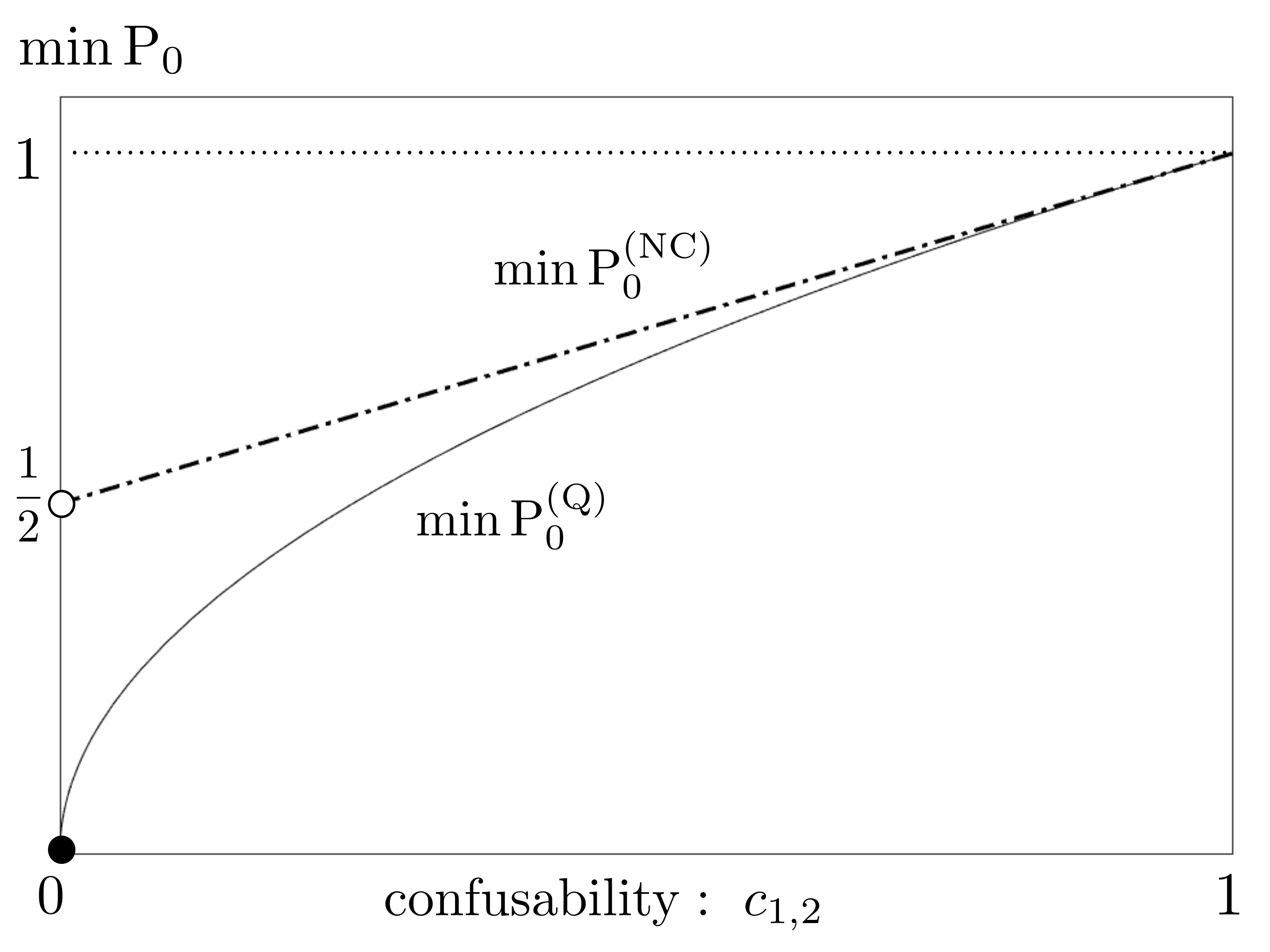}
    \caption{ The minimum value of the inconclusive error rate ${\rm P}_0$ is plotted against the confusability $c_{1,2}>0$ for preparation noncontextual theories (dashed) and quantum theory (sollid). It is shown that the former is strictly greater than the latter, meaning that UD for quantum states contains advantages over a noncontextual model. In a non-contextual theory, the probability of inconclusive outcomes is $0$ for $c_{1,2}=0$. There is a sharp discontinuity, see the main text.}
    \label{fig:udfuns}
\end{figure}

{\it Quantum states}. Let us first consider two pure quantum states $|\psi_1\rangle$ and $|\psi_2\rangle$ for which UD can be performed. The POVM elements may be given as, 
\bea
M_{1} \propto |\overline{\psi}_2\rangle\langle \overline{\psi}_2|~\mathrm{and}~ M_{2} \propto | \overline{\psi}_1\rangle\langle \overline{\psi}_1|
\eea 
where $\langle \overline{\psi}_1 | \psi_1\rangle = \langle \overline{\psi}_2 | \psi_2\rangle =0$. An additional POVM element $M_0$ is needed to collect inconclusive outcomes. The probability of inclusive outcomes for the quantum states denoted by ${\rm P}^{\rm {(Q)}}_{0}$ is known to be \cite{DIEKS1988303, Ivanovic:1987aa, PERES198819}, 
\bea
\min {\rm P}^{\rm (Q)}_{0} = | \langle \psi_1 | \psi_2 \rangle | = \sqrt{c_{1,2}} \label{qmusdrate} 
\eea
where the minimization runs over complete measurements and $c_{1,2}$ is the confusability in Eq. (\ref{eq:confusability}). In Fig. \ref{fig:udfuns}, the probability in Eq. (\ref{qmusdrate}) is plotted.

{\it States with preparation noncontextuality}. We then consider UD for states with preparation noncontextuality. Let us first investigate constraints on response functions $\xi (\lambda)$. Note that a response function corresponding to a sharp measurement can be expressed in the form of Eq. (\ref{eq:xinc}). This can be generalised by including a probabilistic mixture of measurement outcomes. Hence, the most general form of a response function $\xi (\lambda)$ that can be used in unambiguous discrimination will be
\begin{equation} \label{eq:genrf1}
\xi(\lambda) = q \xi_{\y} (\lambda) = \begin{cases} q & \mbox{if } \lambda \in {\rm supp} [\mu_{\y}(\lambda)] \\ 0 & \mbox{if } \lambda \in {\rm supp} [\overline{\mu}_{\y}(\lambda)] \end{cases} 
\end{equation}
for an epistemic state $\mu_{\y}(\lambda)$, which can be freely chosen, and $0\leq q \leq 1$. A response function with the structure above may represent a POVM element in the form  $q | \psi_{\y} \rangle\langle \psi_{\y} |$ in quantum theory. Note that more general response functions could be constructed by mixing multiple outcomes together. However such a response function would be ambiguous. 

The condition that a measurement outcome gives an unambiguous conclusion is
\bea
{\rm P_{M|P}} ( \xi_{\y} | \mu_{\x} ) \propto \delta_{\x,\y}. 
\eea
The condition, applied to a response function in the form of Eq. (\ref{eq:genrf1}), identifies the following response function for two-state UD:
\begin{equation}
\xi_{1}(\lambda) = \begin{cases} q & \mbox{if } \lambda \in {\rm supp} [\overline{\mu}_2 (\lambda)] \\ 0 & \mbox{if } \lambda \in {\rm supp} [\mu_{2} (\lambda)].\end{cases} 
\end{equation}
The same argument applies to the other response function $\xi_{2} (\lambda)$. Note that two states are given with an equal {\it a priori} probability. We can safely assume that the weighting parameter $q\in[0,1]$ remains the same for both response functions $\xi_{1} (\lambda)$ and $\xi_{2} (\lambda)$. The probability of unambiguous outcomes is thus proportional to $q$, which we hence aim to maximise. Equivalently, the probability of inconclusive outcomes is to be minimised.% so this coefficient should take its maximal allowed value.

In fact, two response functions  $\xi_{1} (\lambda)$ and $\xi_{2} (\lambda)$ do not form a complete measurement for the same reason as in UD for quantum states: completeness enforces that $\sum_{\y} \xi_{\y} ( \lambda) = 1$ for all $\lambda \in \Omega$. It is necessary to have an additional response function denoted by $\xi_{0}(\lambda)$ that collects all inconclusive outcomes
\bea
\xi_{0}(\lambda) = 1 - \xi_{1} (\lambda) - \xi_{2} (\lambda).\label{eq:udxi0}
\eea
Note also that $\xi_{0}(\lambda)\geq 0$ for all $\lambda$. The region in which the probability of inconclusive outcomes is minimal can be characterised by the subset 
\bea
\{ \lambda \in \Omega~:~\lambda \in {\rm supp}[ \overline{\mu}_1(\lambda)] \cap {\rm supp}[ \overline{\mu}_2(\lambda)] \} 
\eea 
where both $\xi_1 (\lambda)$ and $\xi_2(\lambda)$ are non-zero. Using the response functions above, it holds that $\xi_{0}(\lambda) = 1-2q$ in the region. Maximising $q$ thus corresponds to minimising the response function $\xi_{0}(\lambda)\geq 0$: one can find $q = 1/2$.

\begin{figure}[]
    \centering
    \includegraphics[scale=0.26]{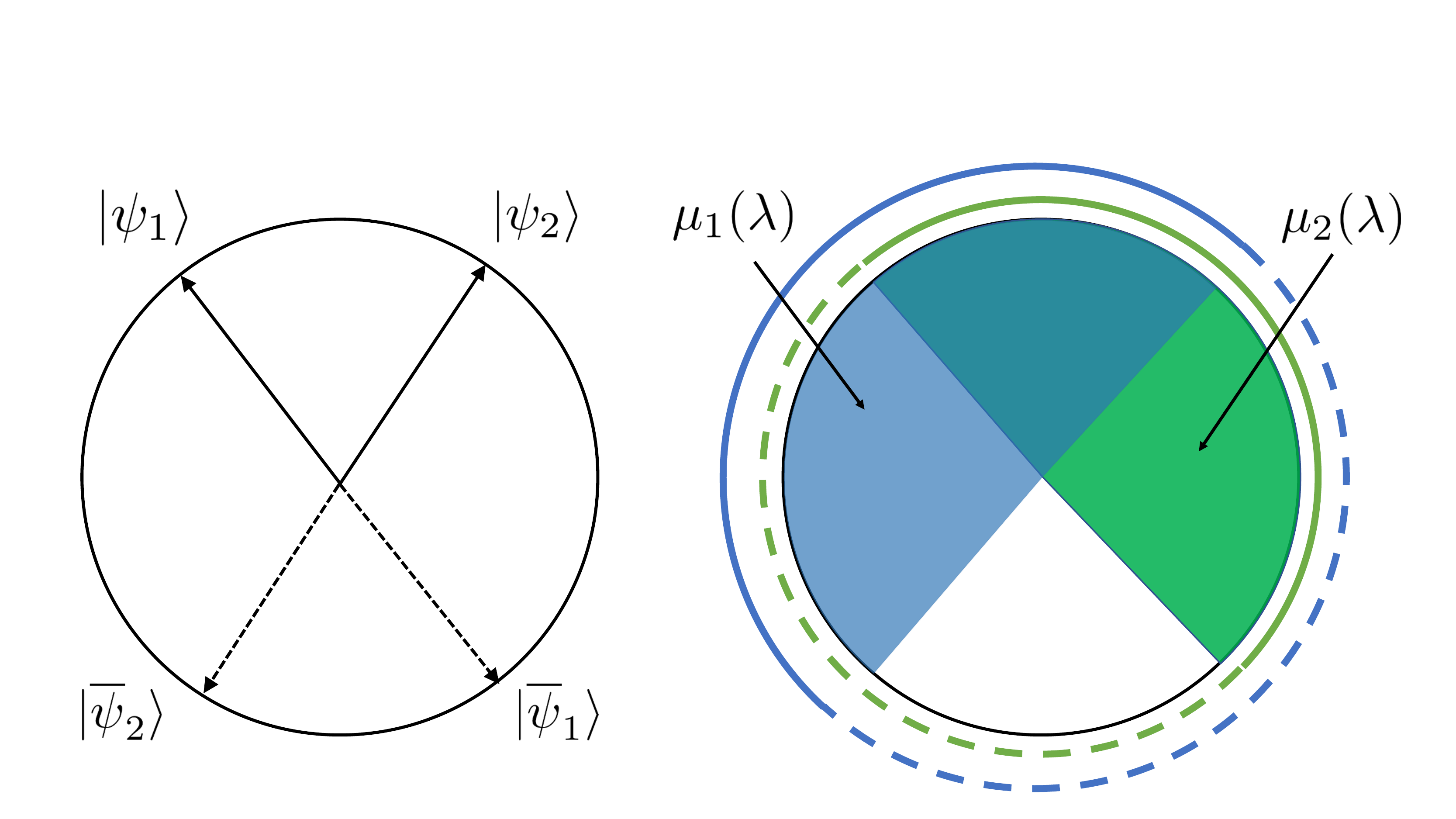}
    \caption{ Two pure states $|\psi_{1}\rangle$ and $|\psi_{2}\rangle$ are considered in UD, for which an optimal measurement corresponds to their orthogonal states $|\overline{\psi}_{1}\rangle$ and $|\overline{\psi}_{1}\rangle$, respectively. The inconclusive outcomes are collected by a POVM element constructed by an equal mixture of given states. The structure is shared with UD of two states ${\mu_1}(\lambda)$ and ${\mu_2}(\lambda)$, diagrammed by blue and green regions respectively. The outer lines signify the supports of reponse functions. $\xi_1(\lambda)$ has the same support as $\bar{\mu}_2 (\lambda)$ (green-dashed line) and $\xi_2(\lambda)$ has the same support as $\bar{\mu}_1 (\lambda)$ (blue-dashed line). }
    \label{fig:udstates}
\end{figure}

The response functions for UD in Eq. (\ref{eq:genrf1}) are thus given by
\bea
\xi_{1}(\lambda) & = & \begin{cases} \frac{1}{2} & \mbox{if } \lambda \in {\rm supp} [\overline{\mu}_2 (\lambda)] \\ 0 & \mbox{if } \lambda \in {\rm supp} [\mu_{2} (\lambda)],\end{cases} \nonumber\\
\xi_{2}(\lambda) & = & \begin{cases} \frac{1}{2} & \mbox{if } \lambda \in {\rm supp} [\overline{\mu}_1 (\lambda)] \\ 0 & \mbox{if } \lambda \in {\rm supp} [\mu_{1} (\lambda)].\end{cases}
\eea
From above and Eq. (\ref{eq:udxi0}), the response function giving inconclusive outcomes can be obtained :
\bea
\xi_{0} (\lambda) &=& \frac{1}{2} \left(1 - 2\xi_{1} (\lambda) \right) + \frac{1}{2} \left(1 -2 \xi_{2} (\lambda) \right) \nonumber \\
&=& \frac{1}{2} \overline{\xi}_1 (\lambda) + \frac{1}{2} \overline{\xi}_2 (\lambda), 
\eea
with $\overline{\xi}_{\y}(\lambda) = 1 - 2 \xi_{\y} (\lambda)$ for $\y=1,2$. Let us write this as
\begin{equation}
\begin{split}
\overline{\xi}_{1}(\lambda) := \xi_{\mu_2} (\lambda) = \begin{cases} 1 & \mbox{if } \lambda \in {\rm supp} [\mu_2 (\lambda)] \\ 0 & \mbox{if } \lambda \in {\rm supp} [\overline{\mu}_{2} (\lambda)],\end{cases} 
\end{split}
\end{equation}
with $\xi_{\mu_2} (\lambda)$ corresponding to a sharp measurement for the epistemic state $\mu_2(\lambda)$. The same argument also applies to the response function $\xi_{\mu_2} (\lambda) $. Bringing all of these together, we have the response function for inconclusive outcomes as follows,
\begin{equation}
    \xi_0 (\lambda) = \frac{1}{2} \xi_{\mu_1} (\lambda) + \frac{1}{2} \xi_{\mu_2}(\lambda). 
\end{equation}
It is therefore shown that the response function is given by a convex combination of two response functions which correspond to sharp measurements for the states in the ensemble. 

In fact, the measurement can be operationally realized by applying two complete sets 
\bea
\{ \xi_{\mu_{1} } (\lambda), \xi_{ \overline{\mu}_{1} } (\lambda) \} ~\mathrm{and}~ \{ \xi_{\mu_{2}} (\lambda), \xi_{ \overline{\mu} _{2} } (\lambda) \}
\eea
with probability $1/2$, respectively. Outcomes with $\xi_{ \overline{\mu}_{\y} } (\lambda)$ for $\y =1,2$ collect inconclusive outcomes and the others lead to unambiguous conclusions. The relevant epistemic states are also depicted in Fig. \ref{fig:udstates} alongside the analogous quantum states.

It is clear that the measurement leads to UD in the following sense. In quantum theory, a measurement strategy of UD consists of three outcomes, two of which show unambiguous detection events and the third of which gives an inconclusive result. In the case of the response functions obtained in a noncontextual theory, there is no chance that the outcome $\overline{\xi}_2$ occurs if $\mu_2 (\lambda)$ is prepared. The epistemic state $\mu_1 (\lambda)$ will likewise never result in the detector associated with the response function $\overline{\xi}_1$ being triggered. These results are, therefore ,unambiguous. The remaining outcomes are $\xi_1(\lambda)$ and $\xi_2 (\lambda)$ and could be triggered by either of the possible epistemic state. These outcomes are collected into the inconclusive outcomes.

Having characterised the optimal measurement, we are now in a position to compute the rate of inconclusive outcomes in noncontextual theories. Given the measurement shown above, the probability of inconclusive outcomes is given as
\bea
\min_{\xi_{0|M}} {\rm P}^{\rm (NC)}_{0} &= &\int_{\Lambda} d\lambda \frac{1}{2}\left (\mu_1(\lambda) + \mu_2(\lambda) \right) \xi_{0|M} (\lambda) \nonumber \\ 
&= &\frac{1}{2} \left( 1 + c_{1, 2} \right).\label{eq:ncud}
\eea
In Fig. \ref{fig:udfuns}, the probabilities of inconclusive outcomes in quantum theory and a contextual model are compared. Hence, contextual advantages for UD of quantum states are shown. 

The caveat is the case when $c_{1,2}=0$, where the rate of inconclusive outcomes in a noncontextual theory is in fact given by $0$. By definition, UD is possible with no inconclusive outcomes. It should be noted that the parameter $q$ in the sharp measurement in Eq. (\ref{eq:genrf1}) can be made equal to one when there is no overlap between the desired response functions. As soon as their supports have some non-zero overlap, no matter how small that region is, the framework enforces that $q \leq 1/2$. There is a discontinuity in the probability of inconclusive outcomes in a noncontextual theory. Therefore, the probability in Eq. (\ref{eq:ncud}) is valid for $c_{1,2}>0$ only. 

%There is a sharp discontinuity between the two cases. 

%but the above equation seems to imply that the ambiguous outcome rate will be $1/2$. The key point is that $q$, our postselection parameter, can be made equal to one when there is no overlap betwen the desired response functions. As soon as their supports overlap, however small that region is, the framework enforces that $q \leq 1/2$. There is a sharp discontinuity between the two cases. 

Finally, it is worth mentioning a physical reason that the aforementioned measurement is a form of UD in a noncontextual theory. There are two classes of measurement possible in quantum theory. Most simply, we can perform projective measurements and probalistically mix the outcomes. Outside of this, we can access a greater set of measurements by entangling the system with an ancilla and then projectively measuring the latter, following the Neumark dilation theorem. An example of this type would be a measurement of the three symmetric qubit states, which requires entanglement with a qutrit. However, as this resource is not available in a noncontextual theory, only the first class can be implemented. Indeed, it has been previously shown that the correlations available to a preparation noncontextual model must be local \cite{PhysRevX.8.011015}. This prevents access to the wider class of POVM elements and we are restricted to the form which was just found.

\subsection{Contextual advantages for MCM}
\label{subsectionIID}

In this subsection, we consider two mixed quantum states for which UD cannot be achieved. Maximum confidence discrimination can be, however, defined, for which we show contextual advantages over a noncontextual model. 

%In this section we show that there are also quantum advantages associated with maximum confidence measurements. To this end, we consider two mixed states for which unambiguous discrimination cannot be performed. MCMs for individual states in quantum and noncontextual models are computed. 

%If the state discrimination ensemble consists of two pure states, the maximum confidence measurement will be unambiguous state discrimination and hence we return to the previous section. To demonstrate MCM in general, we must consider two noisy states and find the maximal confidence for detecting one of them.

\subsubsection{ MCM in quantum theory }

We consider a pair of mixed quantum states given with equal {\it a priori} probabilities,
\bea
\rho_1 & = & (1-p) \ketbra{\psi_1} + p \frac{\I}{2} \nonumber \\
\rho_2 & = &(1-p )\ketbra{\psi_2} + p \frac{\I}{2}.  \label{eq:twons}
\eea
The confusability for two pure states is denoted by $|\ip{\psi_1}{\psi_2}|=\sqrt{c_{1,2}}$. It is straightforward to find an MCM for a quantum state. Following Eq. (\ref{eq:defexp}), we must evaluate
\bea
\max C^{(\Q)} (1)=\max \frac{q_1\tr[ M_1\rho_1]}{\tr[M_1\rho ]} 
\eea
where the maximisation runs over POVM elements and $\rho = (\rho_1 + \rho_2)/2$ denotes the ensemble of given states. The maximisation above can be solved as \cite{PhysRevLett.96.070401}, 
\bea
\max C^{(\Q)}(1)=||\sqrt{\rho}^{-1}q_1\rho_1\sqrt{\rho}^{-1}||_\infty 
\eea
where $||\cdot ||_\infty$ denotes an operator norm. One can find the maximum confidence and write it in terms of the confusability as follows,
\bea
\max C^{(\Q)}(1)=\frac{1}{2} \left( 1+\frac{(1-p)\sqrt{1-c_{1,2}}}{\sqrt{1-(1-p)^2c_{1,2}}} \right). \label{eq:qmcn}
\eea
Note that the noiseless case $p=0$ considering two pure states reproduces UD where the maximum confidence is $1$.

\begin{figure}[]
    \centering
    \includegraphics[scale=0.24]{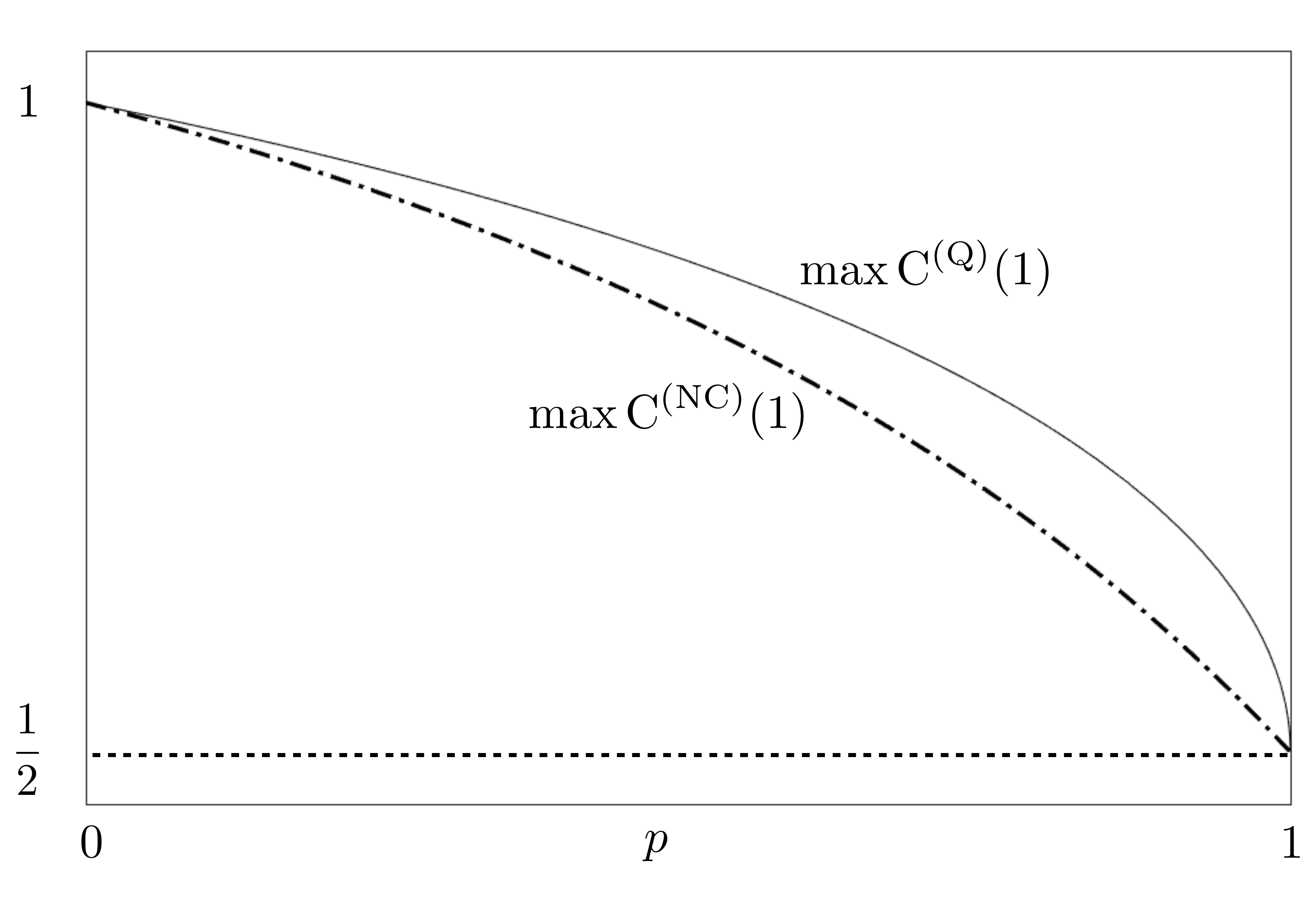}
    \caption{ The maximum confidence is computed for a pair of quantum states (solid) and also in a noncontextual model (dashed). The noise parameter is denoted by $p\in[0,1]$, see Eqs. (\ref{eq:twons}) and (\ref{eq:twonsn}): the case $p=0$ reproduces UD. Contextual advantages for an MCM of noisy quantum states are shown for $p\in (0,1)$. 
    }
    \label{fig:mcmad}
\end{figure}

\subsubsection{ MCM in a noncontextual model}

In the previous subsection \ref{subsectionIIC}, epistemic states $\mu_{\x} (\lambda)$ are associated with pure states $| \psi_{\x} \rangle$ for ${\x}=1,2$. We here consider a noisy preparation in a noncontextual model in the following, 
\bea
\widetilde{ \mu}_1 (\lambda) &=& (1-p) \mu_1(\lambda) + p \mu_{\I/2} (\lambda), \nonumber \\
\widetilde{ \mu}_2 (\lambda) &=& (1-p) \mu_2 (\lambda) + p \mu_{\I/2} (\lambda). \label{eq:twonsn}
\eea
The overall ensemble is then given by
\bea
\mu_P (\lambda) &=& \frac{1}{2}\widetilde{\mu}_1 (\lambda) + \frac{1}{2}\widetilde{\mu}_2 (\lambda) \nonumber \\
&= & p \mu_{\I/2} (\lambda) + (1-p) \left( \frac{1}{2}\mu_1 (\lambda) + \frac{1}{2} \mu_2 (\lambda) \right).~~ \label{eq:noisens}
\eea
The goal is now to compute the maximum confidence, denoted by $\max C^{(\mathrm{NC})}(1)$, for the state $\widetilde{ \mu}_1 (\lambda)$ above, and compare it with the quantum counterpart in Eq. (\ref{eq:qmcn}).

In what follows, let $\xi_{\y}(\lambda)$ denote the response function to find the maximum confidence
%Our aim is to maximise the confidence $C(1)$ associated with a detection of $\mu_1(\lambda)$. We assume that our optimal measurement is sharp and label the response function representing a detection of our chosen state by $\xi_x (\lambda)$, with $x$ denoting the epistemic state which is being sharply measured. The confidence is given by
\begin{equation} \label{eq:conf}
    C^{(\NC)}(1) = \frac{1}{2 \eta_1} \int_{\Lambda} d\lambda \mu_1 (\lambda) \xi_{\y} (\lambda)
\end{equation}
where $\eta_1$ denotes the outcome rate defined by the ensemble and the response function:
\bea \label{eq:outcomewithnoise}
 \eta_1 = \int_{\Lambda} d\lambda \mu_P (\lambda) \xi_{\y} (\lambda). 
\eea
The outcome rate can be rewritten by using Eq. (\ref{eq:noisens}), 
\bea    
\eta_1 &=& 2p \int_{\Lambda} d\lambda \mu_{\I/2} (\lambda) \xi_{\y} (\lambda) \nonumber \\ 
&& + (1-p) \int_\Lambda d\lambda (\mu_1(\lambda) + \mu_2(\lambda) ) \xi_{\y} (\lambda). 
\eea
In a noncontextual theory, it holds that for all $\y$,
\bea
\mu_{\I/2} = \frac{1}{2} \left( \mu_{\y} (\lambda) + \mu_{\bar{\y}} (\lambda) \right)
\eea
which means that the first integral above is equal to $1/2$. The other integral can be expressed in terms of the confusability so that the outcome rate can be written as 
\bea
\eta_1 = p + (1-p) ( c_{1,\y} + c_{2,\y} ). 
\eea
The same argument applies to evaluating the numerator in Eq. (\ref{eq:conf}). After all these steps, we obtain
\bea
 C^{(\NC)}(1) = \frac{1}{2} \left( 1+ \frac{ (1-p) (c_{1,\y} - c_{2,\y} )}{p + (1-p)(c_{1,\y} + c_{2,\y})} \right) \label{eq:Cncmcm}
\eea
which is characterised in terms of the noise parameter $p$ and the confusabilities $c_{1,\y}$ and $c_{2,\y}$. 

It remains to maximise the confidence over response functions. That is, one should maximise the difference between the confusabilities $c_{1,\y}$ and $c_{2,\y}$ while minimising their sum. On the one hand, we recall from MED that the following relation holds  
\bea    
\int_{\Lambda} d\lambda \left( \frac{1}{2} \mu_1 (\lambda) \xi_{\y} (\lambda) + \frac{1}{2} \mu_2(\lambda) \xi_{\bar{\y}}(\lambda) \right) \leq 1 - \frac{c_{1,2}}{2} 
\eea
Note that $\xi_{\bar{\y}} (\lambda) = 1 - \xi_{\y} (\lambda)$ for all $\lambda$. Substituting in this and rearranging then gives us
\begin{equation}
  c_{1,\y} -  c_{2,\y} \leq \left( 1 - c_{1,2} \right), 
\end{equation}
with equality if and only if $\y=\bar{2}$. Thus, the maximum of the difference $c_{1,\y} - c_{2,\y}$ is $1-c_{1,2}$.  

%This can be done by showing that the same $\x$ satisfies both of these conditions.

On the other hand, the sum $c_{1,\y} + c_{2,\y}$ can be bounded from above as follows, 
\bea
        c_{1,\y} + c_{2,\y} &=& 1 + c_{1,\y}  - c_{\bar{2},\y} \nonumber \\
        & \leq & 1 + (1- c_{1,\bar{2}})  \leq 1 + c_{1,2} \nonumber
\eea
It is also bounded from below by
\bea
        c_{1,\y} + c_{2,\y} &=& 2 - c_{\bar{1},\y}  - c_{\bar{2},\y} \nonumber \\
        &\geq & 2 - (1 + c_{\bar{1},\bar{2}})  \geq 1 - c_{1,2}\nonumber
\eea
To summarise, we have shown the upper and lower bounds
\bea
      \left( 1 - c_{1,2} \right) \leq c_{1,\y} +   c_{2,\y} \leq  \left( 1 + c_{1,2} \right). \label{eq:sumbounds}
\eea
Thus, the optimal choice by which the sum $c_{1,\y}$+$c_{2,\y}$ is minimised and also at the same time the difference $c_{1,\y}$ - $c_{2,\y}$ is maximised is given by by $\y=\bar{2}$. We can thus conclude that the maximum confidence in Eq. (\ref{eq:Cncmcm}) is given by the response function $\xi_{\bar{2}}(\lambda)$. Note that the measurement is identical to that in UD. The maximum confidence is then given by
\bea
    \max C^{ (\NC)}(1) = \frac{1}{2} \left( 1 + \frac{(1-p)(1-c_{1,2})}{1- (1-p)c_{1,2}} \right)  \label{eq:ncmcmnoise}
\eea
which is now determined by the noise parameter $p$ and the confusability $c_{1,2}$ only. The case of UD is reproduced by noiseless cases $p=0$. 

%We begin with the relation,
%\bea 
%       c_{1,\x} + c_{2,\x} &=& 1 + c_{1,\x}  - c_{\bar{2},\x} \nonumber \\
%        &\leq& 1 + (1- c_{1,\bar{2}})   \leq 1 + c_{1,2}. \nonumber
%\eea
%This is the upper bound on the sum. 

\begin{figure}[]
    \centering
    \includegraphics[scale=0.35]{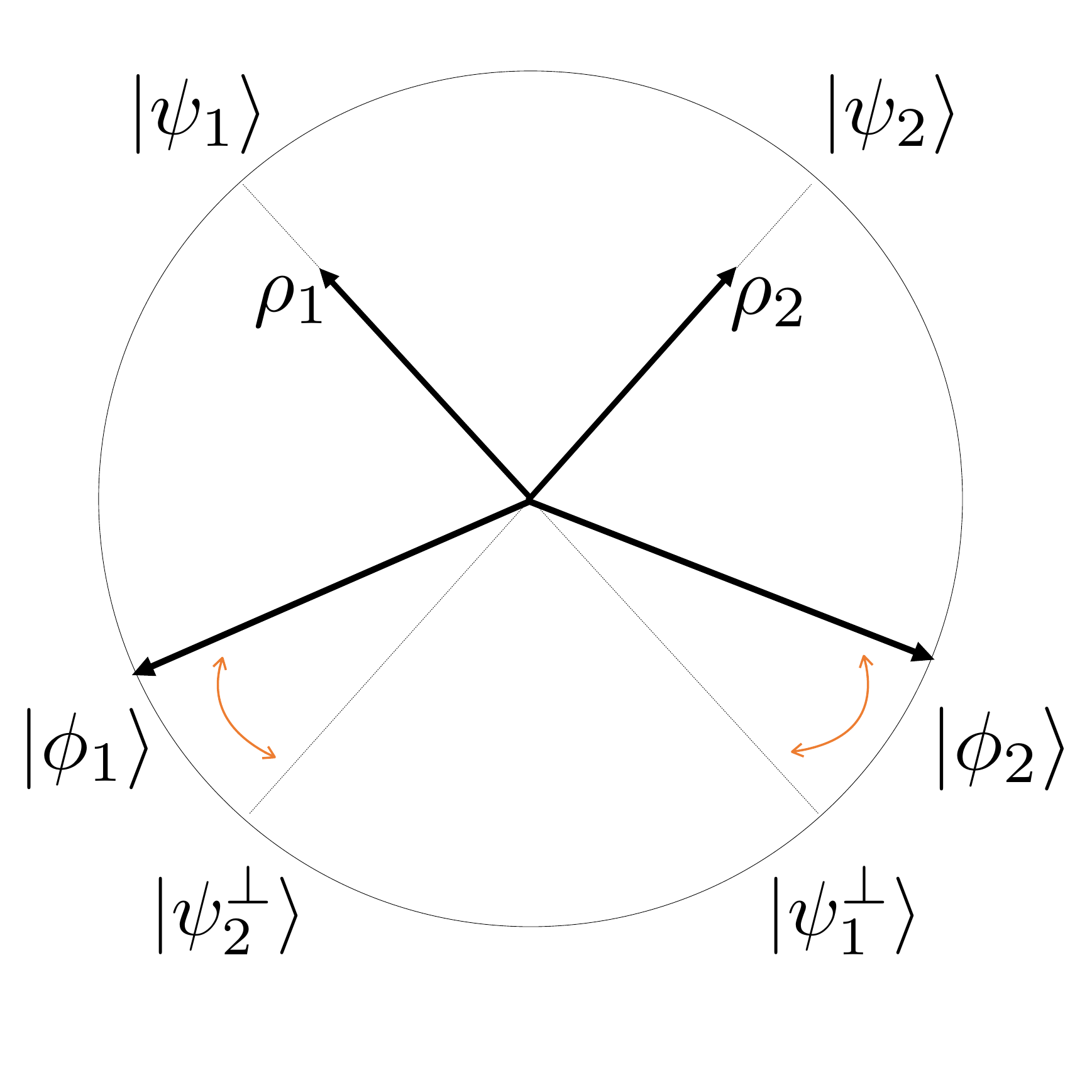}
    \caption{ An MCM for two states $\rho_1$ and $\rho_2$ in Eq. (\ref{eq:twons}) is shown. An optimal POVM element $|\phi_{\y}\rangle \langle \phi_{\y}|$ for state $\rho_{\y}$ for $\y=1,2$ relies on the noise parameter $p$, see Eq. (\ref{eq:optm})}    
    %Measurements for maximum confidence discrimination are compared. For two pure states $|\psi_1\rangle$ and $|\psi_2\rangle$, POVM elements for UD are given by  $|\psi_{1}^{\perp}\rangle$ and  $|\psi_{2}^{\perp}\rangle$. For noisy states $\rho_1$ and $\rho_2$ in Eq. (\ref{eq:twons}), the measurement for maximum confidence discrimination is given by $|\phi_1\rangle$ and $|\phi_2\rangle$. Contrast to the quantum case, a measurement of UD for two states $\mu_1(\lambda)$ and $\mu_1(\lambda)$ in a noncontextual model remains the same for maximum confidence discrimination of their noisy counterpart $\widetilde{\mu}_1(\lambda)$ and $\widetilde{\mu}_1(\lambda)$ in Eq. (\ref{eq:twonsn}) The support of the response function $\xi_1(\lambda)$ is the same as that of $\bar{\mu}_2 (\lambda)$ and the same relation holds for $\xi_2(\lambda)$ and $\bar{\mu}_1 (\lambda)$.  }
    \label{fig:optm}
\end{figure}

\subsubsection{Comparison}

We have computed the maximum confidence for quantum states in Eq. (\ref{eq:qmcn}) and in a noncontextual model in Eq. (\ref{eq:ncmcmnoise}). For $p\in (0,1)$,
\bea
    \max C^{ (\Q)}(1)  >  \max C^{ (\NC)}(1) \label{eq:qbigncmcm}
\eea
holds, which shows contextual advantages for MCMs of quantum states, as seen in Fig. \ref{fig:mcmad}. 

One can investigate maximum confidence measurements in quantum and noncontextual theories. In a noncontextual model, an MCM for the noisy states is identical to the measurement used in UD. This shows that the MCM does not depend on the noise parameter in Eq. (\ref{eq:twonsn}). That is, the measurement realising UD is also an MCM for noisy states in Eq. (\ref{eq:twonsn}). 

Interestingly, an optimal measurement realising UD for two quantum states cannot be extended to noisy states in Eq. (\ref{eq:twons}). Suppose that for the POVM element that performs UD for a state $|\psi_1\rangle$ is given as $M_1\propto |\psi_{2}^{\perp}\rangle \langle \psi_{2}^{\perp}|$. If the measurement is performed on a noisy state $\rho_1$ in Eq. (\ref{eq:twons}), it is not difficult to see that the maximum confidence is equal to Eq. (\ref{eq:ncmcmnoise}) in a noncontextual model. No quantum advantage is concluded. In other words, the noncontextual bound in Eq. (\ref{eq:ncmcmnoise}) can be reproduced in quantum theory by applying the original states' UD measurement to the noisy states. 

In fact, an MCM for the noisy states relies on the noise parameter $p$. To be explicit, an MCM is given by $M_{\y} \propto |\phi_{\y} \rangle \langle \phi_{\y} | $ for $\y = 1,2$ where
\bea
|\phi_{\y} \rangle & = & \sqrt{ \frac{ 1-(1-p) \sqrt{c_{1,2}}}{2} } |0\rangle \nonumber\\
&& + (-1)^{\y + 1} \sqrt{ \frac{ 1 + (1-p) \sqrt{c_{1,2}}}{2} }  |1\rangle  \label{eq:optm}
\eea
for states $\rho_{\y}$, respectively. With the measurement above, the maximum confidence for quantum states in Eq. (\ref{eq:qmcn}) can be obtained, see also Fig. \ref{fig:optm}.

\section{ Certifying maximum confidence  } 
\label{section:cert}

We have so far shown that quantum state discrimination in the forms of MED, UD and MCM generally contains contextual advantages. However, a measurement in a realistic scenario consists of imperfections: it may be neither complete nor sharp. One can therefore ask if the quantum advantages for state discrimination can be obtained in practice when, in particular, undetected events are present. 

In this section, we consider the realistic scenario of quantum state discrimination in a semi-device-independent (sDI) scenario. Namely, a measurement is not yet characterised for an ensemble of quantum states and may also be incomplete. We present a framework for certifying the maximum confidence in the sDI scenario.

 \subsection{Semi device-independent scenario}

Let us begin by presenting the sDI scenario to consider. A set of well-characterised $n$ states, as in Eq. (\ref{eq:ens}), is assumed and detected events are provided. By repeating a prepare-and-measure experiment, the rates of detection events on the $n$ arms are collected. It is also assumed that states are prepared in an {\it independently and identically distributed} manner. The observed probabilities from detectors are denoted by 
\bea
\mathrm{outcome ~rate }:~ \eta_{\mathrm{obs}} = \{\eta_{\y} \}_{\y=0}^n\label{eq:obs}
\eea
where $ \eta_{\y} = \tr[M_{\y} \rho]$ for an ensemble $\rho = \sum_{\x} q_{\x} \rho_{\x}$ and some POVM element $M_{\y}$. Note that $\eta_0$ denotes the collection of undetected events. The probability $\eta_{\y}$ is called an {\it outcome rate} throughout. 

\subsection{ Certification of maximum confidence for quantum states}
\label{subsec:cmcqs}

{\it The framework in quantum theory}. For full generality, we consider an MCM with a predetermined weight $\{ \alpha_{\y}\}_{\y=1}^n$ denoted by
\bea
\langle C^{(\Q)} \rangle_{\alpha} = \sum_{\y=1}^n  \alpha_{\y} C^{(\Q)}(\y). \label{eq:fom}
\eea
The parameters $\{\alpha_{\y} \}$ may define a figure of merit in state discrimination. For instance, if they are identical to the outcome rates, i.e., $\alpha_{\y} = \eta_{\y}$ for $\forall \y\in \{1,\cdots, n \}$, the MCM maximises a success probability in the presence of undetected events. This can be seen in the relation in Eq. (\ref{eq:ba}). When considering an MCM for the $k$-th single detector only, one can put $\alpha_{\x}=\delta_{\x,k}$.

Given an ensemble $S_n$ in Eq. (\ref{eq:ens}) and detected probabilities $\eta_{\mathrm{obs}}$ in Eq. (\ref{eq:obs}), the certification of the maximum confidence is formulated as an optimization problem,
\bea
\mathrm{maximise~} &~&  \langle C^{(\Q)} \rangle_{\alpha} \label{eq:primall} \\
\mathrm{subject~to~} &~& M_{\y}\geq0,~ \sum_{\y=0}^n M_{\y} = \I \nonumber \\
&~& \tr[M_{\y} \rho] = \eta_{\y},~~\y=0,1,\cdots, n \nonumber 
\eea
where $\eta_0$ is the collection of undetected events. The optimisation problem can be solved by a semidefinite program (SDP). This SDP is computationally feasible. Note also that, as it is shown the above, the optimisation problem is equivalent to MED of the $n$ states with {\it a priori} probabilities $\{ \alpha_{\y} q_{\y}/ \eta_{\y}\}_{\y=1}^n$ where a measurement may be incomplete, i.e., $\sum_{\y=1}^n\eta_{\y}<1$.

Similar to MED of quantum states \cite{bae2013}, one can attempt an analytic solution to the optimisation problem in Eq. (\ref{eq:primall}). This can be approached with the linear complementarity problem, which directly considers the optimality conditions. It deals with both the primal and the dual parameters and exploits the general structure lying in the optimisation problem. The primal and the dual problems necessarily give the same result. Since strong duality holds in the optimisation, the optimality can be readily seen from the Karush-Kuhn-Tucker (KKT) conditions. 

\begin{figure}[]
    \centering
    \includegraphics[scale=0.75]{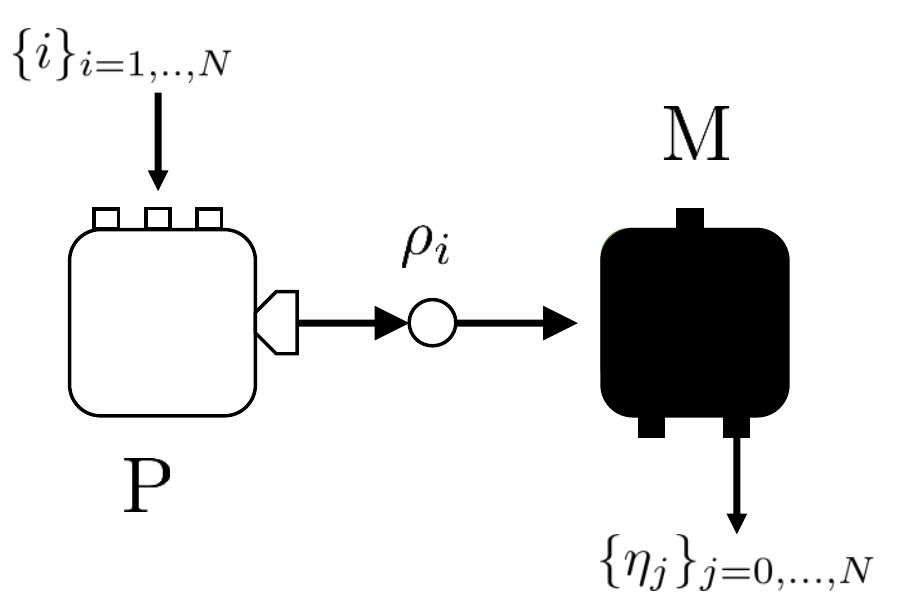}
    \caption{ The prepared quantum states are well-characterised (white). Detectors are arranged to determine which state has been sent. The certification of maximum confidence of a measurement can be obtained from outcome rates from untrusted detectors (black).  }
    \label{fig:example}
\end{figure}

Therefore, the optimality conditions can be summarised as, apart from the constraints in the primal and dual problems, the Lagrangian stability and complementary slackness:
\bea
\mathrm{Lagrangian ~stability} &:&~\forall \y=1,\cdots, n\nonumber \\
 && ~ K = \alpha_{\y} \frac{q_{\y}}{\eta_{\y}}\rho_{\y}+ r_{\y}\sigma_{\y} - s_{\y} \rho, \nonumber\\
&&~  \mathrm{and} ~~ K = r_0 \sigma_0 \label{eq:ls} \\
\mathrm{Complementary~ slackness} &:& ~ \forall \y=0,\cdots, n~~\nonumber \\
 && r_{\y} \tr [ M_{\y} \sigma_{\y} ]=0  \label{eq:cs}
\eea
with dual parameters $K$ and $\{ s_{\y}, r_{\y}, \sigma_{\y} \}$. Note that $\{s_{\y}\}$ and $\{r_{\y}\geq0\}$ are constants and $\{ \sigma_{\y}\}$ quantum states. The Langrangian stability shows the relation between optimal primal and dual parameters. The complementary slackness can be used to find an optimal measurement. 

Once the primal and the dual parameters satisfying the optimality conditions are found, they are automatically optimal and give a solution to the optimisation problem. With the optimal parameters $K^{*}$ and $\{s_{\y}^{*}, r_{\y}^{*}, \sigma_{\y}^{*} \}$ that satisfy the conditions above, the maximum confidence is given as
\bea
\max \langle C^{(\Q)}\rangle_{\alpha} = \tr[K^{*}] + \sum_{\y=1}^n s_{\y}^{*} \eta_{\y}^{*}. 
\eea
A detailed derivation of the optimality conditions is shown in Appendix. \ref{appendix:kkt}. 

{\it Certification of an MCM for a two-state ensemble}. To illustrate the certification scenario, we consider two equally probable states 
\bea
|0\rangle ~\mathrm{and} ~|+\rangle = \frac{1}{\sqrt{2}}(|0\rangle + |1\rangle). 
\eea
Let $C^{(\Q)}(1)$ denote the confidence for the first detector to conclude the state $|0\rangle$. The outcome rate in the first detector is $\eta_1$. From the numerical optimisation in Eq. (\ref{eq:primall}), the certifiable maximum confidence on the first detector is obtained as follows,
\begin{equation*}
\max C^{(\Q)}(1) =
\begin{cases}
1 & \text{if } \eta_{ 1} < 1/4,\\
\in [2/3 , 1] & \text{if }  \eta_{ 1} \in [1/4, 3/4],\\
\in [1/2 ,2/3] & \text{if } \eta_{ 1} > 3/4. 
\end{cases}
\end{equation*}
The maximum confidence above is interpreted as follows. If the outcome rate is low such that $\eta_{ 1} \leq 1/4$, one cannot rule out the possibility that the first detector performs UD. When the outcome rate is more frequent, with $\eta_{ 1} >1/4$, it is clear that the detector cannot perform UD since the maximum confidence is strictly less than $1$. As the outcome rate increases, the maximum confidence on the first arm becomes lower. The example shows a trade-off relation between the maximum confidence and the outcome rate.

\section{ Contextual advantages for certifiable maximum confidence} 
\label{section:adv}

Let us now consider a realistic two-state discrimination scenario in which two states are prepared but three outcomes, including an additional one that collects undetected events, are provided. The certification of the maximum confidence in a detector is investigated and its contextual advantage is analysed. 

\subsection{ Quantum state discrimination in practice}

We here investigate the certifiable maximum confidence in a realistic two-state discrimination in detail. The framework developed in subsection \ref{subsec:cmcqs} is applied to certify the maximum confidence on a single detector. We recall that a pair of two pure states can always be identified by a single parameter $\theta$ such that
\bea
\cos\theta = \langle \psi_1 | \psi_2\rangle = \sqrt{c_{1,2}} 
\eea
with the confusability $c_{1,2}$. This also means that any two-state discrimination problem can be mapped onto a two-dimensional plane spanned by the two states, i.e., $\mathrm{span}\{ |\psi_1\rangle, |\psi_{1}^{\perp}\rangle\} = \mathrm{span}\{ |\psi_2\rangle, |\psi_{2}^{\perp}\rangle\}$. Hence, without loss of generality, a two-state discrimination problem can be safely restricted to a qubit space. Let us write down the ensemble as
\bea
&& |\psi_{  1}\rangle = \cos \frac{\theta}{2} |0\rangle + \sin\frac{\theta}{2} |1\rangle~\mathrm{and} \nonumber\\
&& |\psi_{ 2 }\rangle = \cos \frac{\theta}{2} |0\rangle - \sin\frac{\theta}{2} |1\rangle, \label{eq:twoensemble} 
\eea
which may be prepared with {\it a priori} probabilities $q_1$ and $q_2$, respectively. 

Two detectors are arranged to find which of the states has been sent. A ``click" in the first  detector concludes that the state $|\psi_1\rangle$ was prepared and a detection event in the second one is for the state $|\psi_2\rangle$. The experiment is performed repeatedly so that one finds the rate of detection events in each arm. There are also cases where no detections are reported due to either the loss of prepared quantum systems during transmission or the failure of detectors to respond. After a measurement is repeated, outcome rates are found to be 
\bea
\eta_{\mathrm{obs}} = \{ \eta_0, \eta_1,\eta_2\},   
\eea 
where $\eta_0$ is the rate of undetected events.

For outcome rates compatible with quantum theory, there exist POVM elements $\{ M_{\y}\}_{\y=1}^2$ for the ensemble $\rho$ such that
\bea
\eta_{\y} & = & \tr[M_{\y} \rho]~ ~\mathrm{where}  ~~\rho =  \sum_{\x=1,2} q_{\x}  |\psi_{\x}\rangle \langle \psi_{\x}|
\eea
where the measurement fulfills the condition, $M_0 + M_1 + M_2 = \I$. In general, the figure of merit can be written for predetermined parameters $\alpha = \{ \alpha_1, \alpha_2\}$,
\bea
\max  \langle C^{(\Q)} \rangle_{\alpha} = \max ~\left( \alpha_1 C^{(\Q)}(1) + \alpha_2 C^{(\Q)}(2) \right) 
\eea
where the maximization runs over POVM elements. The optimisation problem can be solved analytically with the optimality conditions in Eqs. (\ref{eq:ls}) and (\ref{eq:cs}). 
 
\begin{figure}[]
    \centering
    \includegraphics[scale=0.25]{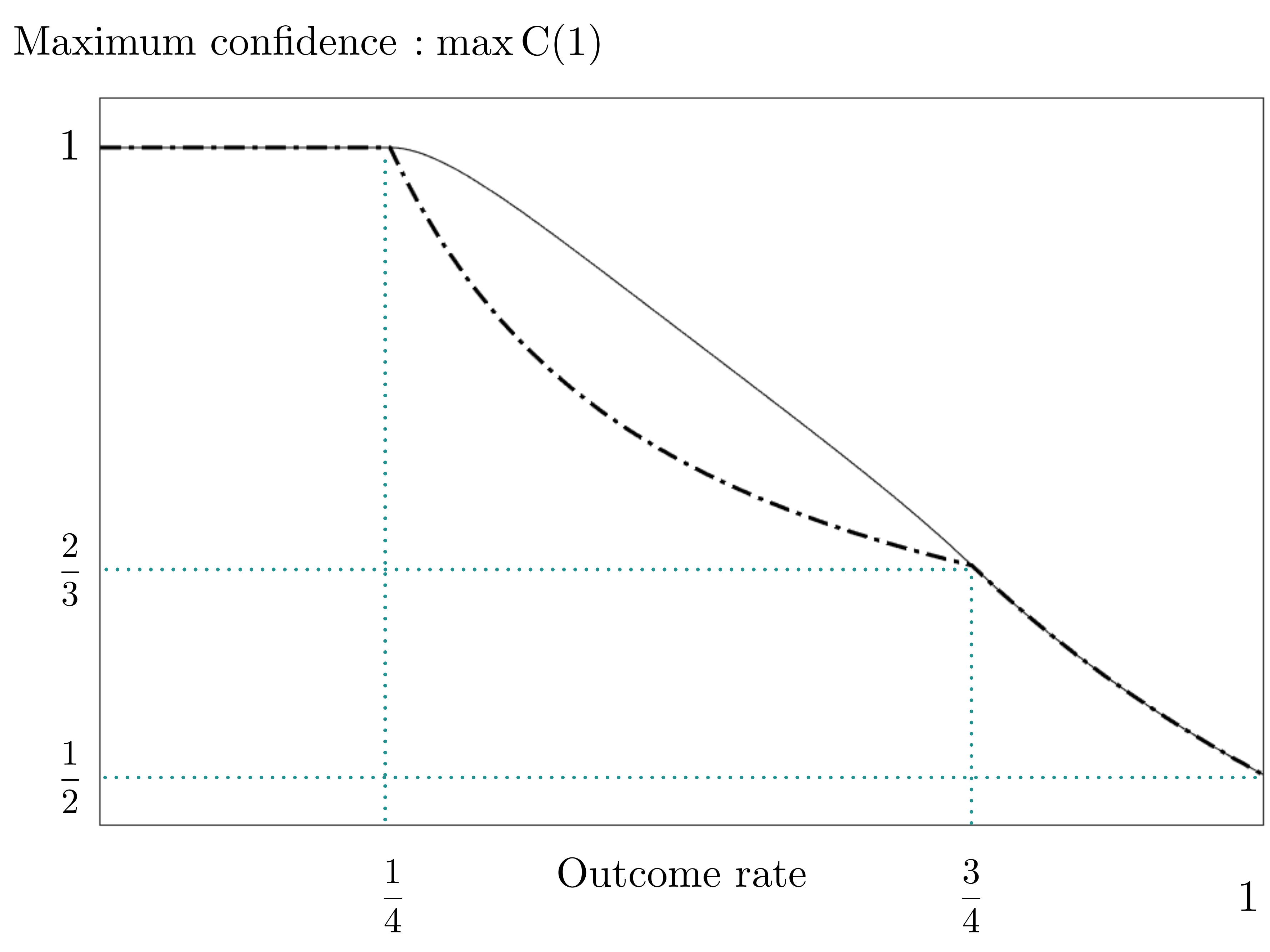}
    \caption{ In two-state discrimination between $|0\rangle$ and $(|0\rangle + |1\rangle)/\sqrt{2}$, the certifiable MCM in the first detector, denoted by $\max C(1)$, is plotted with respect to outcome rate $\eta_1$. The certifiable MCM is shown in both quantum theory (solid line) and in a noncontextual ontological model (dotted line). A low detection rate $\eta_1\leq 1/4$ is compatible with UD. The contextual advantages exist whenever an outcome rate is within the range $\eta_1\leq 3/4$. However, no contextual advantage can be obtained if an outcome rate is too high for $\eta_1>3/4$.       }
    \label{fig:my_label}
\end{figure}

\subsection{ Maximum confidence on a quantum state} 
\label{subsec:consdetector}

The maximum confidence in the realistic two-state discrimination scenario above can be certified as follows. For simplicity, let us assume the preparation of equiprobable states, i.e., $q_1=q_2=1/2$ and show the certification for the first detector. The detailed derivation is shown in Appendix \ref{appendix:certifiedquantumconfidence}.

In the certification scenario, a detector in a two-state discrimination scenario shows an outcome rate $\eta_1$ when the measurement is repeated. Using our KKT conditions, it can be shown that the certifiable maximum confidence on such a measurement is given by 
\bea
&& \max C^{(\Q)} (1) = \nonumber \\ \label{quantumrates}
&& \begin{cases}
~ 1, & \mathrm{for}~\eta_1 \in [0, c_{-}]  \\ 
~ \frac{1}{2}+ f(\eta_1, c_{1,2})  & \mathrm{for}~ \eta_1 \in [ c_{-},  c_{+}] \\
~ \frac{1}{2\eta_1}, & \mathrm{for}~ [ c_{+}, 1]   
\end{cases}
\eea
where
\bea
c_{\pm} & = &\frac{1}{2} (1\pm c_{1,2} )~~\mathrm{and} \nonumber \\
f( \eta_1, c_{1,2}) & = & \frac{1}{4\eta_1} \sqrt{ \left( \frac{1-c_{1,2}}{c_{1,2}} \right) \left(   c_{1,2} - (1-2\eta_1)^2 \right) }. \nonumber
\eea
 Note that certification depends upon the outcome rate $\eta_1$ of detected events only for a given ensemble of states.

An optimal measurement for maximum confidence discrimination can be characterized according to the outcome rate. For an outcome rate $\eta_1  \leq c_+$, an optimal measurement is given by rank-one POVM elements. For $\eta_1 > c_+$ where the outcome rate is relatively higher, the maximum confidence is obtained from a rank-two POVM element. One can find that that too frequent detection events, i.e., $\eta_1\geq c_+$, rule out a rank-one measurement for maximum confidence discrimination: thus, a rank-two measurement is also certified.

%In this case, the maximum confidence is given by $1/(2\eta_1)$. The rate of detection events can, in fact, be constrained by certifiable MCM.  

\subsection{Contextual advantage }  
\label{subsectionc}

We now investigate the certification of an MCM in a noncontextual theory and compare it with the quantum case. To this end, the main task is to optimise a measurement in a noncontextual ontological model, i.e., a response function $\xi_1(\lambda)$ in the first arm, given the extra constraint with a fixed outcome rate $\eta_1$. We write two epistemic states as $\mu_1 (\lambda)$ and $\mu_2(\lambda)$, which show the confusability $c_{1,2}$ that is the same as that of quantum states defined in Eq. (\ref{eq:twoensemble}). %$\xi_1(\lambda)$ identifies $\mu_1(\lambda)$.

The fixed outcome rate must first be addressed. The outcome rate can be expressed in terms of the confusabilities as 
\begin{equation}
    \eta_1 = \frac{1}{2} c_{1,\y} + \frac{1}{2} c_{2,\y}.
\end{equation}
where $\y$ labels the sharp response function for the epistemic state $\mu_{\y}(\lambda)$. We can see, following Eq. (\ref{eq:sumbounds}), that a sharp measurement will only be able to attain outcome rates in the range
\begin{equation}
    c_- \leq \eta_1 \leq  c_+.
\end{equation}
For rates less than the lower bound, we must use a sharp measurement weighted by a probability. Such response functions were seen in Eq. (\ref{eq:genrf1}). For rates above this bound, a ``rank-2" response function (i.e., one consisting of mixing multiple outcomes) is required. We note that these boundaries are exactly the same as those from the quantum case, see Eq. (\ref{quantumrates}). Each region of our piecewise confidence function will be addressed in what follows.

Let us begin with the infrequent detection region where $\eta_1 \leq c_-$. Here we must again use a response function of the form
\begin{equation} \label{eq:genrf}
\xi_{1} (\lambda) = q \xi_{\y} (\lambda) = \begin{cases} q & \mbox{if } \lambda \in {\rm supp} [\mu_{\y}(\lambda)] \\ 0 & \mbox{if } \lambda \in {\rm supp} [\overline{\mu}_{\y}(\lambda)] . \end{cases} 
\end{equation}
With this function we can express the confidence as 
\begin{equation} \label{eq:ncconf}
    \begin{split}
        C^{(\NC)} (1) &= \frac{q}{2\eta_1} \int_{\Lambda} d\lambda \mu_{1}(\lambda) \xi_{\y} (\lambda) = \frac{q }{2\eta_1}c_{1,\y}.
    \end{split}
\end{equation}
The goal is to maximise the confusability over a constant outcome rate. To take the latter into account, we use the $\ell_1$ distance \cite{Lostaglio2020contextualadvantage}. In a noncontextual theory, this is related to the confusability as follows
\bea
c_{\x,\y} &=& 1- \frac{1}{2} || \mu_{\x} - \mu_{\y} ||_1,\label{conl1} \\
\mathrm{where} && || \mu_{\x} - \mu_{\y} ||_1 = \int_{\Omega} d\lambda |\mu_{\x} (\lambda) - \mu_{\y} (\lambda) |.\nonumber
\eea
We now express $\eta_1$ in terms of the $\ell_1$ distance:
\begin{equation}
\begin{split}
\eta_{1}  &= \frac{q}{2} \left( c_{\y, 1} + c_{\y,2} \right) \\
&= \frac{q}{2} \left( 2 - \frac{1}{2} \| \mu_{\y} - \mu_{1} \|_1 - \frac{1}{2} \| \mu_{\y} - \mu_{2} \|_1 \right). \label{effdis}
\end{split}
\end{equation}
The triangle inequality allows us to exploit the relation, 
\bea \label{eq:triangleinequality}
\|  \mu_{\y} - \mu_2 \|_1 \leq \| \mu_{\y} - \mu_{1} \|_1 + \| \mu_{1} - \mu_{2} \|_1. 
\eea
Combining this result with Eq. (\ref{conl1}) above and writing in terms of $c_{1,2}$, we obtain
\bea
\| \mu_{\y}  - \mu_{1} \|_1 \geq 1 - 2 \frac{\eta_1}{q} + c_{1,2},
\eea
or, in a more convenient form using the confusabilities,
\bea
c_{1,\y} \leq \frac{\eta_1}{q} + c_{-}.  \label{l1inequality}
\eea
Bringing all of these together, the maximum confidence can be expressed as 
\bea
\max C^{(\NC)} (1) = \max_q \left( \frac{1}{2}  + \frac{ c_{-} }{2  \eta_1} q\right), 
\eea
where the maximisation runs over the variable $q \in[0,1]$.  

Let us now find the certified maximum confidence given an outcome rate $\eta_1$. For the range of the outcome rate where $\eta_1< c_-$, the optimal parameter can be chosen as $q = 2 \eta_1 / c_-$. Thus, the certifiable maximum confidence is given as $C^{(\NC)}(1)=1$. The cases can be interpreted as UD, except that the confidence of the detector's other arm is not yet specified. Therefore, a distinction between the quantum and noncontextual theories is not found in terms of the maximum confidence of a given state. Of course, as it has been shown above, there is a distinction in terms of a different figure of merit, the rate of ambiguous outcomes.

The next range to consider is when the outcome rate is within the bounds, $\eta_1 \in [ c_-, c_+]$ where we recall $c_{\pm} = (1 \pm c_{1,2})/2$. Here, sharp measurements give the desired outcome rate and, therefore, are treated simply by letting $q=1$ in the above calculation. This gives a maximum confidence,
\bea
\max C^{(\NC)} (1) = \frac{1}{2}\left( 1 + \frac{1-c_{1,2}}{2\eta_1} \right)< \max C^{(\Q)} (1),~~~~~~~\label{eq:ncc}
\eea
with the maximum confidence in quantum theory in Eq. (\ref{quantumrates}). Thus, a quantum advantage is shown in the range, see Fig. \ref{fig:my_label}.

%\begin{figure*}[]
 %   \centering
  %  \includegraphics[scale=0.45]{respfun}
   % \caption{ Series of response functions (solid colours) satisfying the maximum confidence available to preparation noncontextual theories. The ensemble consists of epistemic states $\mu_1 (\lambda)$ (horizontal lines) and $\mu_2(\lambda)$ (vertical lines). The hashed region signifies the overlap of the two epistemic states. The function changes with $\eta_1$, increasing in magnitude between $\eta_1=0$ and $\eta_1 = (1 -c_{1,2})/2$, then rotating around the ontic state space, then occupying a greater region which also increases in magnitude until, at $\eta_1 =1$ occupying the entire space.      }
%    \label{fig:respfun}
% \end{figure*}

For the high-outcome-rate range where $\eta_1 \geq c_+$, we deduce the response function by considering the behaviour at two values of $\eta_{1}$. The confidence must be continuous at the point $\eta_{1} = c_+$ and the response function at this point is a sharp measurement of $\mu_1(\lambda)$. The response function for higher values of $\eta_1$ must consist of binning together multiple measurement outcomes due to the bounds on sharp measurements. At $\eta_1=1$, the response function will be equal to one across the whole ontic state space, which can be decomposed into a sum of two non-overlapping sharp measurements. We can see that the response function will take the form
\begin{equation} \label{eq:rank2resp}
    \xi_1 (\lambda) = \xi_{\mu_1}(\lambda) + a \xi_{\overline{\mu}_1} (\lambda)
\end{equation}
where $a$ is some constant that can be determined by evaluating the associated outcome rate. Doing this gives
\bea
 \xi_1 (\lambda) = \xi_{\mu_1}(\lambda) + \frac{\eta_1 - c_+}{1-c_+} \xi_{\overline{\mu}_1} (\lambda). \label{eq:resfunction}
\eea
This function gives a outcome rate $\eta_{1}$ and a maximum confidence 
\bea
\max C^{(\NC)} (1) = \frac{1}{2\eta_1}
\eea
which is again the same as the behaviour in the quantum case.

Let us summarise the key features of the response function, which is optimized according to the outcome rate $\eta_1\in [0,1]$. In the low outcome region with $\eta_1\leq c_-$, the optimal response function has the same support as the state $\mu_2 (\lambda)$, on which it linearly increases from zero to one as the outcome rate goes from zero to $c_{-}$. In the central region with $\eta_1\in[c_-, c_+]$, the optimal response function corresponds to a projective measurement which slightly shifts its support away from $\mathrm{supp} [ \mu_2 (\lambda)]$ and towards $\mathrm{supp} [\mu_1(\lambda)]$, to which it coincides when $\eta_1 = c_{+}$. Finally, when the outcome rate is even higher for $\eta_1\geq c_+$, the support includes the rest of the ontic state space. The response function increases linearly on the region of two supports $\mathrm{supp} [ \xi_1 (\lambda)]$ and $ \mathrm{supp} [ \overline{\mu}_1(\lambda) ]$. When $\eta_1=1$, the response function will be equal to one for all ontic states.

%We can summarise the key features of the response function as it changes between $\eta_1=0$ and $\eta_1=1$. In the low outcome region, it has the same support as the state $\mu_2 (\lambda)$ and, in this region, has a uniform strength which increases from zero to one as the outcome rate goes from zero to $c_{-}$. In the central region it is a projective measurement which 'rotates' its support away from ${\rm supp}\[\mu_2(\lambda\]$ and towards ${\rm supp}\[\mu_1(\lambda\]$, which it reaches when $\eta_1 = c_{1}$. Finally, when the outcome rate is high, the support includes also the rest of the ontic state space, and, on the region where ${\rm supp} [\xi_1 (\lambda)] \cup {\rm supp}\[\overline{mu}_1(\lambda\]$, will increase in strength again from zero to one. When $\eta_1=1$, the response function will be equal to one for all ontic states.

Interestingly, the three ranges showing distinct forms of the response functions in a noncontextual model and an optimal measurement in quantum theory precisely coincide with each other. Contextual advantages in terms of a higher maximum confidence are shown in the central region only, where a sharp measurement turns out to be optimal, see also Fig. \ref{fig:my_label}. In the next section, noisy preparations are considered where the aforementioned properties do not hold in general. Contextual advantages in terms of a higher maximum confidence appear over the whole range of outcome rates. The ranges giving distinct forms of a measurement in quantum and noncontextual theories no longer coincide with each other.

\section{Certifiable maximum confidence on noisy preparation}
\label{section:noisyadv}

We consider a noisy preparation and investigate contextual advantages in the certification of an MCM. We first recall the result in subsection \ref{subsectionIID} that the contextual advantages for the MCM hold true for noisy quantum states. We here extend the contextual advantage to the certification scenario. Again, let us consider a pair of mixed states given with equal {\it a priori} probabilities
\bea
\rho_1 & = & (1-p) \ketbra{\psi_1} + p \frac{\mathbb{I}}{2},~\mathrm{and} \\
\rho_2 & = &(1-p )\ketbra{\psi_2} + p \frac{\I}{2}. \label{eq:twoqn}
\eea
We also exploit the confusability for the pure states, $c_{1,2}=|\langle \psi_1 | \psi_2 \rangle |^2$. In what follows, we compute the certified maximum confidence when the outcome rate is given by $\eta_1$ in the first arm. 

\subsection{Quantum states}

We apply the same method used in Section \ref{section:cert} to compute the certifiable maximum confidence. The detailed derivation is shown in in Appendix \ref{appendix:certifiedquantumconfidence}. It is fairly straightforward to obtain the results. Contrary to the noiseless case in Section \ref{section:cert}, it is found that the ranges in which different kinds of measurements are optimal do not coincide between quantum and noncontextual theories. The certifiable maximum confidence can be summarised depending on the range of the outcome rate. 

Firstly, when the outcome rate is in the range $\eta_1 \in [0,  \eta_{1}^{(-)}]$ where 
\bea
\eta_{1}^{(\pm)}=\frac{1}{2} \left(1 \pm (1-p)^2 c_{1,2} \right), 
\eea
the confidence is given by,
\bea
\max C^{(\Q)}(1) = \frac{1}{2} \left(1+\frac{ (1-p) \sqrt{1-c_{1,2}}}{\sqrt{1 - (1-p)^2 c_{1,2} }} \right). 
\eea
Note that the noiseless case $p=0$ reproduces UD and also the boundary condition in the range $\eta_{1}^{(-)} = c_-$ in Eq. (\ref{quantumrates}). For noisy cases with $p>0$, it holds that  $\eta_{1}^{(-)} > c_-$.

Secondly, when $\eta_1 \in [\eta_{1}^{(-)},\eta_{1}^{(+)}]$ the certifiable maximum confidence is computed as
\bea
\max C^{(\Q)} (1)&=& \frac{1}{2} + g_p( \eta_1,c_{1,2}) 
%(1+ \sqrt{\frac{1-c_{1,2}}{c_{1,2}}} \times \nonumber \\
%&& \sqrt{\frac{ (1-p)^2 c_{1,2} - ( 1 - \frac{1}{2 \eta_1}(1- (1-p)^2 c_{1,2}))^2}{1- (1-p)^2 c_{1,2}}}) \nonumber
\eea
where 
\bea
 g_p( \eta_1,c_{1,2} ) = \frac{1}{4\eta_1}\sqrt{ \left(\frac{1-c_{1,2}}{c_{1,2}} \right) \left( (1-p)^2c_{1,2} -(1-2\eta_1)^2 \right) }.  \nonumber
\eea
Note that the case $p=0$ reproduces the certifiable maximum confidence in a noiseless case in Eq. (\ref{quantumrates}).

Thirdly, when the outcome rate is in the range $\eta_1 \in [\eta_{1}^{(+)},1]$, the certifiable maximum confidence is obtained as
\bea
\max C^{(\Q)} (1) = \frac{1}{2} \left( 1+\frac{ (1-p) \sqrt{1-c_{1,2}}}{\sqrt{1-(1-p)^2 c_{1,2} }}\left(\frac{1}{\eta_1}-1 \right) \right). 
\eea
Note that it holds that $\eta_{1}^{(+)}<c_+$ for noisy cases with $p>0$, see Eq. (\ref{quantumrates}). In addition, the noiseless case $p=0$ also reproduces Eq. (\ref{quantumrates}).

\subsection{Noncontextual model}

Similarly to what is shown in subsection \ref{subsectionIID}, we consider noisy states $\tilde{\mu}_1 (\lambda)$ and $\tilde{\mu}_2 (\lambda)$ as defined in Eq. (\ref{eq:twonsn}) with {\it a priori} probabilities $1/2$, respectively. In the certification scenario, it is assumed that the outcome rate in the first arm is given by $\eta_1$. We then aim to find the certifiable maximum confidence on, say, the first arm. 

\begin{figure}[]
    \centering
    \includegraphics[scale=0.25]{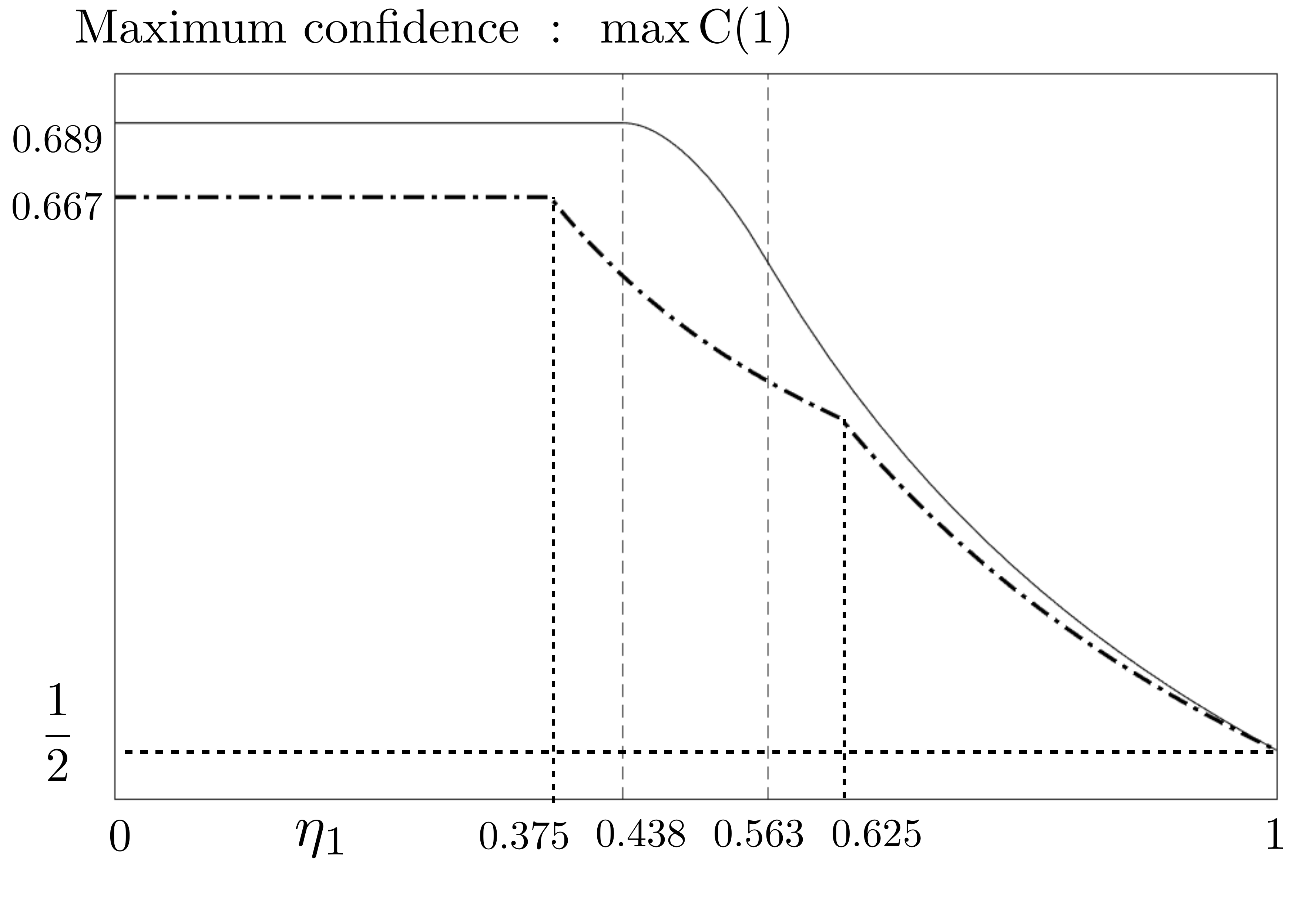}
    \caption{ The certifiable maximum confidence is shown for two noisy states $c_{1,2}=1/2$ and $p=1/2$ where $p$ is the noise parameter in Eq. (\ref{eq:twoqn}). The certifiable maximum confidence varies depending on a outcome rate $\eta_1$. The certifiable maximum confidence for quantum states (sollid) is higher than that in a noncontextual model (dotted) for all $\eta_1\in [0,1]$. }
    \label{fig:figcad}
\end{figure}

Sharp measurements cannot reproduce all outcome rates, as shown in Eq. (\ref{eq:sumbounds}). The outcome rate $\eta_1$ can be obtained using 
\bea
\eta_1 = \int_{\Lambda} d\lambda \mu_P (\lambda) \xi_{\x} (\lambda) 
\eea
where $\mu_{P}(\lambda)$ denotes the ensemble in Eq. (\ref{eq:noisens}). Applying Eq. (\ref{eq:sumbounds}) to the integral above, one can obtain bounds on the outcome rate as follows,
\begin{equation}
    \frac{1}{2} \left( 1 - (1-p) c_{1,2})\right) \leq \eta_1 \leq  \frac{1}{2} \left( 1 + (1-p) c_{1,2}\right). \label{eq:range}
\end{equation}
Note that a sharp measurement can produce the desired statistics in the range above. If we require an outcome rate is below the lower bound, we must use weighted sharp measurements. If the desired outcome rate is higher than the upper bound, we must use rank-2 equivalent measurements. Interestingly, the boundaries in a noncontextual theory are different from those in the quantum case in the previous section (see also Fig. \ref{fig:figcad}). 

Once the outcome rate is in the range $\eta_1 \in [0, (1-(1-p)c_{1,2})/2]$, the measurement for the maximum confidence must be a weighted sharp measurement, i.e., we again let $\xi_1 (\lambda) = q \xi_{\y} (\lambda)$ where $0 \leq q \leq 1$ and $\xi_{\y}(\lambda)$ is a sharp measurement for a to-be-determined epistemic state.

For this response function, it holds that
\bea
    \eta_1 = \frac{q}{2} \left( p + (1-p) (c_{1,\y} + c_{2,\y}) \right)  
\eea
which can be written as,
\bea
    q = \frac{2\eta_1}{p + (1-p)(c_{1,\y}+c_{2,\y})}.
\eea
Thus, the value $q$ is obtained from a given $\eta_1$. Let us express the confidence in terms of the confusabilities,
\bea
        C^{(\NC)}(1) &=& \frac{1}{2\eta_1} \int_{\Lambda} d\lambda \tilde{\mu}_1 (\lambda) \xi_1 (\lambda) \nonumber \\
        &=& \frac{q}{2\eta_1} \int_{\Lambda} d\lambda \tilde{\mu}_1 (\lambda) \xi_{\y} (\lambda) \nonumber \\
        &=& 1 - \frac{p + 2(1-p)c_{2,\y}}{p + (1-p)(c_{1,\y}+c_{2,\y})}.
\eea
To find the maximum confidence, one has to minimise the fraction by finding $\y$ such that the numerator is minimal and the denominator is maximal. It turns out that the optimal choice is given by $\y=\bar{2}$. It is obvious that $c_{2,\y}$ is minimized with $\y=\bar{2}$. From Eq. (\ref{eq:sumbounds}), the sum $c_{1,\y} + c_{2,\y}$ is minimal as $1-c_{1,2}$. Therefore, we have
\bea
    \max C^{(\NC)}(1) = 1 - \frac{p}{2\left( 1 - (1-p)c_{1,2} \right)}
\eea
which also shows that the noiseless case $p=0$ reproduces the case UD. 

When the outcome rate is in the range in Eq. (\ref{eq:range}), the measurement must be sharp and we again use $\xi_1 (\lambda) = \xi_{\y} (\lambda)$ to avoid confusion between response functions. We apply the same technique used in subsection \ref{subsectionc}. The key tool is the inequality, 
\bea
    1 - c_{\y,2} \leq 2 - c_{1,\y} - c_{1,2},
\eea
which follows from the triangle inequality Eq. \ref{eq:triangleinequality}. Note also that 
\bea 
   \eta_1 = \frac{p}{2} + \frac{1-p}{2} \left( c_{1,\y} + c_{2,\y} \right), 
\eea
from which,
\bea
    c_{2,\y} = \frac{2\eta_1 - p}{1-p} - c_{1,\y}. 
\eea
All these imply that 
\bea
    c_{1,\y} \leq \frac{1}{2} \left( 1 + \frac{2 \eta_1 -p}{1-p} - c_{1,2} \right). 
\eea
The confidence is given by
\bea
    C^{(\NC)}(1) = \frac{1}{2\eta_1} \left( \frac{p}{2} + (1-p) c_{1,x} \right), 
\eea
which has the maximum as follows,
\bea
    \max C^{(\NC)}(1) = \frac{1}{2} + \frac{(1-p)(1-c_{1,2})}{4\eta_1}. 
\eea
This agrees with Eq. (\ref{eq:ncc}) when $p=0$.

Again, in the range $\eta_1 \in [(1+(1-p)c_{1,2})/2 , 1]$ when the outcome rate is high, the response function can be directly deduced. The respones function will take the form
\bea
    \xi_1 (\lambda) = \xi_{\mu_1}(\lambda) + a \xi_{\bar{\mu}_1} (\lambda) 
\eea
as in Eq. \ref{eq:rank2resp} and for the same reasons, where $\xi_{\mu_1}(\lambda)$ is the sharp measurement associated with $\mu_1(\lambda)$. Note that the value $a$ is fixed by the outcome rate and can be found by calculating the $\eta_1$ given by the response function,
\bea
    a = \frac{2 \eta_1 - 1 - (1-p)c_{1,2}}{1 - (1-p)c_{1,2}}. 
\eea
The confidence is therefore obtained as
\bea
    \max C^{(\NC)}(1) = \frac{1}{2\eta_1} \left(1 - \frac{p(1-\eta_1)}{1 - (1-p)c_{1,2}} \right). 
\eea
This agrees with Eq. (\ref{eq:resfunction}) for cases $p=0,1$. 

%We can now compare the two types of theory to see when a quantum advantage is possible... [still to be filled in]

\subsection{Comparison}

In both the noiseless and noisy cases, in subsections \ref{section:adv} and \ref{section:noisyadv}, it is seen that the maximum confidence can be characterised into three ranges of low, intermediate, and high outcome rates. The feature commonly shared between them is that the maximum confidence does not increase as the outcome rate gets more frequent: a less frequent outcome rate implies a higher the maximum confidence and vice versa.  

Contrasting the cases, it is shown that the ranges characterising the maximum confidence coincide in quantum and noncontextual theories when the preparation is noiseless. Contextual advantages are shown in the intermediate range only. In the noisy case, the ranges are distinct in quantum and noncontextual theories, where the intermediate range becomes narrower. Contextual advantages in this scenario appear in the whole range of outcomes rates. 

It is observed that the contextual advantages appearing in the low and high outcome rates are related with each other. Let us consider the range of lower detection rate in a noisy case,
\bea
\eta_1 <     \frac{1}{2} \left( 1 - (1-p) c_{1,2})\right). 
\eea
The gap between quantum and noncontextual theories is denoted by, 
\bea
\Delta_L := \max C^{(\Q)} (1)- \max C^{(\NC)}(1). 
\eea
One can find that the gap is strictly positive if $p>0$ and zero for $p=0$. Then, for a higher outcome rate where
\bea
\eta_1 >    \frac{1}{2} \left( 1 + (1-p) c_{1,2})\right)
\eea
it turns out that the gap between quantum and noncontextual theories can be written as,
\bea
\Delta_H: = \max C^{(\Q) } (1)- \max C^{(\NC) } (1)=\left( \frac{1}{\eta_1} - 1 \right)\Delta_L. 
\eea
which is also strictly positive for $p>0$. If no contextual advantage appears in the low-outcome-rate range, i.e., $\Delta_L=0$, then neither does it when the outcome rate is high, i.e., $\Delta_H=0$.

\section{Conclusion}
\label{section:conc}

State discrimination is a fundamental tool in information processing in general. Its central role in quantum information applications motivates us to investigate exactly when quantum theory provides an advantage compared to classical theories. 

In the present contribution, we have demonstrated contextual advantages in the general setting of maximum confidence measurements, which includes minimum error and unambiguous discrimination as particular cases. We have extended the contextual advantages of quantum state discrimination to the cases of UD and maximum confidence discrimination. Note that an MCM presents a unifying general framework of state discrimination such as UD and MED. We also examine the optimal measurement. It turns out that an MCM in a noncontextual theory remains identical in the presence of a uniform noise. However, an MCM in quantum theory varies according to the ratio of noise: it depends upon how much noise is present in given states. Consequently, an MCM for noisy states shows a higher maximum confidence compared with a noncontextual theory. 

Having found that the contextual advantages for state discrimination exist in general, we show how to certify the maximum confidence in a realistic scenario, where the outcome rates are provided for an ensemble of states while a measurement is not fully characterised. Note also that undetected events may be present. Along the way, an optimisation problem is introduced for the certification of the maximum confidence on quantum states. The certification of the maximum confidence in a noncontextual ontological model is developed and then compared with quantum cases. It turns out that one can always find contextual advantages in the certification of the maximum confidence on quantum states. Our results show how quantum state discrimination can achieve its advantages over a noncontextual ontological model. 

While our work has generalised two-state discrimination to a wider range of figures of merit, there is still much room to generalise further by considering a wider class of ensembles. In particular, three-state discrimination poses an interesting problem due to the impossibility of creating the symmetric three-state ensemble in a noncontextual theory \cite{PhysRevA.71.052108}. Exploring such areas will further our understanding of the quantum-classical boundaries.

Our results set the ground for understanding how quantum information applications that exploit quantum state discrimination can achieve advantages over a classical theory in a realistic scenario. Among the tasks using state discrimination, it would be interesting to investigate randomness generation, e.g., \cite{PhysRevApplied.7.054018}. It would also be interesting to investigate contextual advantages in quantum computing tasks, such as quantum machine learning, where state discrimination is often processed to manipulate classical data over the limitations of conventional computing \cite{lloyd2020quantum}.

\section*{Acknowledgement}
KF, HL, and JB were supported by National Research Foundation of Korea (NRF-2021R1A2C2006309), Institute of Information \& communications Technology Planning \& Evaluation (IITP) grant (Grant No. 2019-0-00831, the ITRC Program/IITP-2021-2018-0-01402). JBB and CRC were supported by the Independent Research Fund Denmark and a KAIST-DTU Alliance stipend.

\bibliography{referencecol}
\bibliographystyle{apsrev4-1}

%\begin{thebibliography}{99}

%\end{thebibliography}

\newpage 

\appendix

\section{Derivation of the optimality condition in the certification scenario } \label{appendix:kkt}

\begin{widetext}

We here derive the optimality conditions in Eqs. (\ref{eq:ls}) and (\ref{eq:cs}), which allow for the certification of an MCM given specified outcome statistics. That is, given outcome rates $\eta_{\y}$, the goal is to maximise $\sum_{\y} \alpha_{\y} C_{\y}$ over a measurement, where
\bea
C_{\y}=\frac{q_{\y}}{\eta_{\y}}\tr[M_{\y} \rho_{\y}].\nonumber
\eea 
In fact, the optimisation problem can be written as an SDP. The primal problem is the following:
\bea
p^* = \max  && \sum_{\y=1}^n \alpha_{\y} \frac{q_{\y}}{\eta_{\y}} \tr[ M_{\y} \rho_{\y} ] \nonumber\\
\mathrm{subject~to}&~~& M_{\y} \geq 0, ~\sum_{ {\y}=1}^n M_{\y} \leq I, \nonumber\\
&& \tr[\rho M_{\y} ] = \eta_{\y} \nonumber
\eea
Let $M_0 =I - \sum_{{\y}= 1}^n M_{\y} \geq 0$ denote a slack variable that takes undetected events into account. Let us introduce dual variables $r_{\y} \sigma_{\y}$ for inequality constraint where $r_{\y} \geq 0$ and $\sigma_{\y}$ is a quantum state, and $K$ and $s_{\y}$ to derive the Lagrangian functional in the following,
\bea
&~~&\mathcal{L} (\{M_{\y} \}_{{\y}=0}^n, \{r_{\y}\}_{{\y}=0}^n, \{\sigma_{\y}\}_{{\y}=0}^n, \{s_{\y}\}_{{\y}=1}^n, K) \nonumber \\
&~~&=\sum_{{\y}=1}^n \alpha_{\y} \frac{q_{\y}}{\eta_{\y}}\tr[M_{\y} \rho_{\y}]+\sum_{{\y} = 0}^n r_{\y}\tr[M_{\y}\sigma_{\y}]+
+\tr[K(I-\sum_{{\y}=0}^n M_{\y})]+\sum_{{\y}=1}^n s_{\y} (\eta_{\y}- \tr[\rho M_{\y}]) \nonumber \\
&~~&=\sum_{{\y}=1}^n \tr[M_{\y} (\alpha_{\y} \frac{\rho_{\y}}{\eta_{\y}}+r_{\y}\sigma_{\y}-K - s_{\y} \rho)]+\tr[M_0(r_0\sigma_0-K)]+\tr[K]+\sum_{{\y}=1}^n s_{\y} \eta_{\y}. \nonumber
\eea
The dual functional is derived as follows, 
\bea
&~~&g(\{r_{\y}\}_{{\y} = 0}^n,\{\sigma_{\y}\}_{ {\y}=0}^n,\{s_{\y}\}_{{\y}=1}^n,K) \nonumber\\
&~~&=\sup_{ \{ M_{\y} \}_{\y=0}^n }\mathcal{L}(\{M_{\y}\}_{{\y}=0}^n, \{r_{\y}\}_{{\y}=0}^n, \{\sigma_{\y}\}_{{\y}=0}^n, \{s_{\y}\}_{{\y}=1}^n, K)\nonumber \\
&~~&=\begin{cases}  \tr[K]+\sum_{{\y}=1}^n s_{\y} \eta_{\y} & \mbox{if }  \frac{\alpha_{\y} }{\eta_{\y}} \rho_{\y}+ r_{\y}\sigma_{\y} - K- s_{\y}\rho=0 \mbox{ and } r_0 \sigma_0 - K=0 , ~~{\y}=1,2,\cdots,n \nonumber \\
+\infty & \mbox{otherwise. } \nonumber
\end{cases}.
\eea
Since the dual functional does not diverge, we have that
\bea
K = r_0 \sigma_0,~\mathrm{and}~K = \alpha_{\y} \frac{\rho_{\y}}{\eta_{\y}}+r_{\y}\sigma_{\y} - s_{\y} \rho,~~{\y}=1,2,\cdots n.\nonumber
\eea
This condition is called the \textit{Lagrangian stability}. The dual problem can be written as,
\bea
d^* = \min && \tr[K]+\sum_{{\y}=1}^n s_{\y} \eta_{\y} \nonumber \\
\mathrm{subject~to}&~~& K+s_{\y}\rho\geq \frac{\alpha_{\y}}{\eta_{\y}}\rho_{\y},~\mathrm{and}  \nonumber\\
 && K\geq 0. \nonumber
\eea
In general, it holds that $p^{*}\geq d^{*}$. The equality holds when the problem is strictly feasible. For instance, one can choose $M_{\y} = \eta_{\y}I$ for all $\y$ to show that the primal problem is strictly feasible. We thus have that $p^* = d^*$. 

When the dual and primal problems give the same solution, one can also solve the optimisation problem by analyzing the optimality conditions directly. For the SDP above, the optimality conditions are listed as,
\bea
    \text{(Lagrangian stability) }  K & = & \alpha_{\y} \frac{q_{\y}}{\eta_{\y}}\rho_{\x}+ r_{\y}\sigma_{\y} - s_{\y}\rho,~~ \forall \y \nonumber\\\
    K & = & r_0\sigma_0 \nonumber \\
    \text{ (Complementary slackness) } r_{\y} \tr[M_{\y} \sigma_{\y}] &=& 0, ~~\forall \y
\eea
together with the constraints in the primal and dual problems. Although the optimality conditions contain a greater number of variables than the primal and dual problems, they are useful for exploiting the generic structure existing in an optimisation problem.

\section{Solving the optimality conditions for certifying the maximum confidence } %\label{appendix:mcqnoisy}
\label{appendix:certifiedquantumconfidence}

We here show the approach of the so-called linear complementarity problem in the certification of a maximum confidence. We consider qubit states and show how the optimality conditions can be directly analysed. 

Suppose that two states $\rho_1$ and $\rho_2$ are given with {\it a priori} probability $1/2$, respectively, 
\bea
\rho_1 & = & (1-p) \ketbra{\psi_1} + p \frac{\mathbb{I}}{2},~\mathrm{and} \\
\rho_2 & = &(1-p )\ketbra{\psi_2} + p \frac{\I}{2} \nonumber
\eea
for which the outcome rates given by $\eta_1$ and $\eta_2$. The goal is now to find the certifiable maximum confidence on the first arm. Let us begin with the following primal problem:
\bea
p^* = \max &~&\frac{1}{2\eta_1} \tr[ M_1 \rho_1 ] \nonumber\\
\mathrm{subject~to} &~~& 0 \leq M_1 \leq I, \nonumber\\
&& \tr[ M_1 \rho ] = \eta_1. \nonumber
\eea
%Introduce $X_1,X_2\geq 0$ dual variables for inequality constraints and $\lambda$ for equality constraint. 
The Lagrangian function can be constructed as
\beas
\mathcal{L}(M_1,X_1,X_2,\lambda) &~&=\frac{1}{2\eta_1}\tr[\rho_1 M_1]+\tr[X_1 M_1]+\tr[(I-M_1)X_2]+\lambda(\eta_1-\tr[\rho M_1])\\
&~&=\lambda \eta_1+\tr[X_2]+\tr[(\frac{\rho_1}{2\eta_1}+X_1-X_2-\lambda \rho)M_1]
\eeas
from which the dual problem can be obtained:
%\beas
%g(X_1,X_2,\lambda)&~& =\sup_{M_1} \mathcal{L}(M_1,X_1,X_2,\lambda)\\
%&~& =\begin{cases} \lambda\eta_1+\tr[X_2] & \mbox{if } \frac{\rho_1}{2\eta_1}+X_1-X_2-\lambda \rho=0 \\ +\infty & \mbox{otherwise. }
%\end{cases}.
%\eeas
%the dual problem is defined as
\bea
d^* = \min &&  \lambda \eta_1+\tr[X_2] \nonumber\\
\mathrm{subject~to}&~~& X_2+\lambda \rho \geq \frac{1}{2\eta_1}\rho_1, \nonumber \\
&& X_2\geq 0 \nonumber
\eea

The optimality conditions can be found and listed out as follows,
\bea
X_1-X_2 &=&\lambda \rho-\frac{1}{2\eta_1}\rho_1 \label{eq:ab1}\\
X_1,X_2 &\geq& 0 \nonumber \\ 
M_1 X_1  &=& 0\nonumber \\
(I-M_1)X_2 &=&0\nonumber \\
0 \leq &M_1&\leq I\nonumber \\
\tr[\rho M_1]&=&\eta_1.\nonumber 
\eea
Since qubit measurements are considered, $X_1 X_2=0$ holds. Since the non-negative operators $X_1$ and $X_2$ are orthogonal, they can be obtained from the spectral decomposition in Eq. (\ref{eq:ab1}). Let $\nu_\pm$ denote the positive and negative eigenvalues $\ket{\nu_\pm}$, respectively, so that 
\beas
X_1-X_2&=&\lambda\rho-\frac{1}{2\eta_1}\rho_1 = \nu_+\ketbra{\nu_+}+\nu_-\ketbra{\nu_-}
\eeas
where
\beas
\nu_\pm&=&\frac{\tan\theta}{4\eta_1}(\gamma\pm\sqrt{1+\gamma^2}\bar{p}\cos\theta)\\
\ket{\nu_\pm}&=&\frac{1}{\sqrt{2+2\gamma^2\mp 2\gamma\sqrt{1+\gamma^2} }}(\ket{0}+(\gamma\mp \sqrt{1+\gamma^2})\ket{1})
\eeas
with $\gamma=(2\eta_1\lambda-1)\cot\theta$ and $\bar{p}=1-p$. It is straightforward to find the maximum confidence,
\beas
\max C^{(\Q)}(1)&=&\lambda \eta_1+\tr[X_2]\\
&=&\frac{1}{2}(1+\gamma\tan\theta)-\nu_-\\
&=&\frac{1}{2}+\frac{\tan\theta}{4\eta_1}[(2\eta_1-1)\gamma+\bar{p}\cos\theta\sqrt{1+\gamma^2}],
\eeas
where the parameter $\gamma$, relying on the dual parameter $\lambda$, needs to be further optimised. If either $X_1$ or $X_2$ is of full-rank, then the optimisation becomes trivial since $M_1=0$ or $I$. Assuming $X_1$ and $X_2$ are not full-rank, there are three possible cases for $X_1 X_2=0$. 

Firstly, we consider that $X_1=0$ and $X_2 >0$. Since $X_1=0$, we have that $\nu_+=0$,
\beas
\gamma=\frac{-\bar{p} \cos\theta}{\sqrt{1- \bar{p}^2\cos^2\theta}}
\eeas
and
\bea
\max C^{(\Q)}(1) & = &\frac{1}{2} \left[1+\frac{\bar{p} \sin\theta}{\sqrt{1-\bar{p}^2\cos^2\theta}}\left(\frac{1}{\eta_1}-1\right) \right], \nonumber \\
~\mathrm{where}~~~\ket{\nu_-}& = &\frac{1}{\sqrt{2}}(\sqrt{1+\bar{p} \cos\theta}\ket{0}+\sqrt{1 -  \bar{p} \cos\theta}\ket{1}). \nonumber
\eea
Since $X_2=-\nu_-\ketbra{\nu_-}$ and $X_2(I-M_1)=0$, 
\beas
M_1&=&\ketbra{\nu_-}+\alpha \ketbra{\nu_+} = \alpha I + (1-\alpha) \ketbra{\nu_-}
\eeas
for some constant $0\leq \alpha\leq 1$. That is, the optimal measurement is a convex combination of $I$ and $\ketbra{\nu_-}$. To find $\alpha$, we use condition $\tr[ M_1\rho]=\eta_1$ so that 
\beas
\alpha=\frac{2\eta_1-1-\bar{p}^2\cos^2\theta}{1-\bar{p}^2\cos^2\theta}.
\eeas
Since $\alpha \geq 0$, the outcome rate is constrained by $\eta_1\geq \frac{1}{2}(1+\bar{p}^2\cos^2\theta)$.

Secondly, we consider that $X_2=0$ and $X_1>0$. For $X_2=0$, for which it holds that $\nu_-=0$. 
\beas
\gamma=\frac{ \bar{p} \cos\theta}{\sqrt{1- \bar{p}^2\cos^2\theta}}.
\eeas
Since $X_1$ is rank-one, the optimal measurement must be rank-one $M_1=\beta \ketbra{\nu_-}$. It is straightforward to find the maximum confidence,
\bea
\max C^{(\Q)}(1)& = &\frac{1}{2}(1+\frac{ \bar{p} \sin\theta}{\sqrt{1- \bar{p}^2\cos^2\theta}}) \nonumber \\
\mathrm{where}~~\ket{\nu_-} &=& \frac{1}{\sqrt{2}}(\sqrt{1- \bar{p} \cos\theta}\ket{0}+\sqrt{1 +  \bar{p} \cos\theta}\ket{1}).  \nonumber
\eea
The optimal measurement is given by
\beas
M_1=\frac{2\eta_1}{1- \bar{p}^2\cos^2\theta} \ketbra{\nu_-}.
\eeas
The condition $\beta\leq1$ is equivalent to $\eta_1\leq \frac{1}{2}(1-\bar{p}^2\cos^2\theta)$.

Thirdly, $X_1>0$ and $X_2>0$. Since $X_1$ and $X_2$ are both rank-one, optimal POVM elements $M_1$ and $I-M_1$ are also rank-one so that $M_1=\ketbra{\nu_-}$. From the condition $\tr[ M_1\rho ]=\eta_1$, we find
\bea
\gamma=\frac{1-2\eta_1}{\sqrt{ \bar{p}^2\cos^2\theta-(1-2\eta_1)^2}}. \nonumber
\eea
We then have,
\bea
\max C^{(\Q)}(1) &=& \frac{1}{2}+\frac{\tan\theta}{4\eta_1}\sqrt{ \bar{p}^2\cos^2\theta-(1-2\eta_1)^2} \nonumber \\
\mathrm{where}~~\ket{\nu_-}& =&\frac{1}{\sqrt{2}} \left(\sqrt{1-\frac{1-2\eta_1}{ \bar{p} \cos\theta}} ~|0\rangle+\sqrt{1+\frac{1-2\eta_1}{ \bar{p} \cos\theta}}~ \ket{1}\right). \nonumber
\eea
The conditions $\nu_+\geq0$ and $\nu_- \leq 0$ are equivalent to $\frac{1}{2}(1 - \bar{p}^2\cos^2\theta)\leq \eta_1 \leq \frac{1}{2}(1+\bar{p}^2\cos^2\theta)$. 

\end{widetext}

\end{document}